%% file: MAIN/main.tex
\documentclass[letterpaper]{MAIN/nature}

\usepackage{color}
\usepackage{url}
\usepackage{verbatim}
\usepackage{multirow}
\usepackage{xspace}
\usepackage{graphicx}
\usepackage{amsmath} 
\usepackage{bbm}
\usepackage{amssymb}
\usepackage{booktabs} 
\usepackage{times} 
\usepackage{lscape}
\usepackage[left=1in,right=1in,bottom=1.1in,top=1in]{geometry}
\usepackage{enumitem}
\usepackage[table,xcdraw]{xcolor}
\usepackage{xcolor}
\usepackage[normalem]{ulem}
\usepackage[innercaption]{sidecap}
\usepackage{lineno}
\usepackage{stmaryrd}
\usepackage{xr-hyper}
\usepackage{xurl}
\usepackage[htt]{hyphenat}
\usepackage{makecell}
\usepackage{tcolorbox}
\usepackage{siunitx}

\usepackage[colorlinks,citecolor=blue,urlcolor=magenta]{hyperref}
\makeatletter
\newcommand*{\addFileDependency}[1]{
  \typeout{(#1)}
  \@addtofilelist{#1}
  \IfFileExists{#1}{}{\typeout{No file #1.}}
}
\makeatother

\usepackage[colorlinks,citecolor=blue,urlcolor=magenta]{hyperref}
\usepackage[margin=-1cm, labelfont=bf, font=footnotesize]{caption}
\usepackage{algorithm}
\usepackage{algorithmic}

\newcommand{\SIref}[1]{{\color{red}#1}}

\newcommand{\hide}[1]{}
\newcommand{\xhdr}[1]{\vspace{1em}\noindent{{\bf #1}}}

\newif\ifshowrevisions
\showrevisionsfalse 

\ifshowrevisions
  \definecolor{revisionbg}{HTML}{FFF8D6}
  \definecolor{revisionline}{HTML}{C9AA45}
\else
  \definecolor{revisionbg}{HTML}{FFFFFF}
  \definecolor{revisionline}{HTML}{FFFFFF}
\fi

\AtBeginDocument{%
  \NewCommandCopy\revorigcite\cite
  \NewCommandCopy\revorigref\ref
}

\ifshowrevisions
  \newcommand{\revision}[1]{%
    \bgroup
    \renewcommand{\cite}[1]{\mbox{\revorigcite{##1}}}%
    \renewcommand{\ref}[1]{\mbox{\revorigref{##1}}}%
    \color{black}%
    \markoverwith{%
      \smash{%
        \rlap{\textcolor{revisionbg}{%
          \rule[-0.6ex]{2pt}{2.8ex}}}%
        \textcolor{revisionline}{%
          \rule[-0.6ex]{2pt}{0.4pt}}%
      }%
    }%
    \ULon{#1}%
  }
\else
  \newcommand{\revision}[1]{#1}
\fi

\usepackage{comment}
\usepackage{bm}
\graphicspath{{./FIG/}} 

\title{\begin{center}Clinical Note Bloat Reduction for Efficient LLM Use\end{center}}

\author  
{\begin{center}   
Jordan L. Cahoon$^{1,2,*}$, Chloe Stanwyck M.D., M.S.$^{1,3,*}$\textbf{}, Asad Aali M.S.$^{4}$, Rachel Madding M.D.$^{5}$, Sulaiman S. Somani M.D.$^{6}$, Emma Sun M.S.$^{7}$, Yixing Jiang$^{1}$, Renumathy Dhanasekaran M.D.$^{8}$, and Emily Alsentzer Ph.D.$^{1,7,9,\ddag}$ \\[1mm]  
\small{$^{1}$Department of Biomedical Data Science, Stanford University, Stanford, CA} \\
\small{$^{2}$Department of Pathology, Stanford University, Stanford,
CA} \\
\small{$^{3}$Department of Anesthesiology, Perioperative and Pain Medicine, Stanford University, Stanford,
CA} \\
\small{$^{4}$Department of Radiology, Stanford University, Stanford,
CA} \\
\small{$^{5}$Department of Obstetrics and Gynecology, Stanford University, Stanford,
CA} \\
\small{$^{6}$Department of Medicine, Stanford University, Stanford, CA} \\
\small{$^{7}$Department of Computer Science, Stanford University, Stanford, CA} \\
\small{$^{8}$Division of Gastroenterology and
Hepatology, Stanford University, Stanford,
CA} \\
\small{$^{9}$Weill Cancer Hub West} \\
[1mm]
\small{$\ddag$Corresponding author. Email: ealsentzer@stanford.edu}\\
\small{$^*$Equal contribution} 
\end{center}
}

\begin{document}

\maketitle

{\spacing{1.4}

\section*{Abstract}

\begin{abstract}
 \input{MAIN/000abstract}
\end{abstract}
}

\clearpage

\spacing{1.38}

\section*{Introduction}

\input{MAIN/010intro}

\section*{Methods}

\input{MAIN/040methods}

\section*{Results}

\input{MAIN/020results}

\section*{Discussion}

\input{MAIN/030discuss}



\clearpage


 


\section*{Acknowledgments} 
This work is supported in part by the Weill Cancer Hub West Initiative and a Stanford HAI-Accenture Research Grant. 
 J.L.C. is supported by the Warren Alpert Computational Biology \& Artificial Intelligence Fellowship. C.S. is supported by the National Institute of General Medical Sciences of the National Institutes of Health under award number T32GM089626. A.A. is supported by NIH grant R01 HL167974 and ARPA-H contract AY2AX000045. This work is supported in part by the Weill Cancer Hub West Initiative. 
 Some of the computing for this project was performed on the Stanford Carina cluster. We would like to thank Stanford University and Stanford Research Computing for providing computational resources and support that contributed to these research results, as well as the Technology \& Digital Solutions Team for assisting with data access. 


\section*{References}

{
\spacing{0.85}

\bibliographystyle{MAIN/naturemag}
\bibliography{MAIN/refs}
}

\input{MAIN/050figures}

\clearpage
\newgeometry{left=0.9in,right=0.9in,top=1in,bottom=1in}

\setcounter{figure}{0}
\setcounter{table}{0}
\renewcommand{\figurename}{Supplementary Figure}
\renewcommand{\tablename}{Supplementary Table}

\renewcommand{\theHfigure}{supp.\arabic{figure}}
\renewcommand{\theHtable}{supp.\arabic{table}}

\captionsetup{margin=0pt,labelfont=bf,font=footnotesize}
\setlength{\tabcolsep}{3pt}
\spacing{1}

\begin{center}
{\Large Supplementary Information}
\end{center}

\clearpage
\section*{Supplementary Notes}
\input{SI/SI-notes}

\clearpage
\section*{Supplementary Figures and Tables}
\input{SI/SI-figures}

\clearpage

\end{document}

%% file: MAIN/000abstract.tex

\begin{abstract}
\normalfont

\noindent\textbf{Background:} Health systems are rapidly deploying large language models (LLMs) that use clinical notes for decision support. However, modern documentation practices rely heavily on templates, copy–paste shortcuts, and auto-populated fields that produce extensive duplicated text (``note bloat''). These documentation artifacts dilute clinically meaningful signal, limit usable longitudinal context, and substantially increase the computational cost of LLM-based applications.

\noindent\textbf{Methods:} 
We introduce TRACE, a scalable pre-processing pipeline that removes note bloat by leveraging note-level EHR metadata to identify templated and copied content and applying frequency-based de-duplication when metadata are unavailable. 
We evaluate TRACE using blinded physician span review and gold-standard templated-text annotations. We apply TRACE to four real-world cohorts spanning liver transplant, obstetrics, and inpatient populations across multiple \revision{health systems}, totaling $\approx$ 5.3M notes. We compare zero-shot LLMs and embedding-based classifiers on original vs TRACE-processed notes for information extraction (1\revision{3} liver transplant tasks; 7 obstetrics tasks) and outcome prediction (5-year overall survival, postpartum hemorrhage, and 30-day readmission).

\noindent\textbf{Results:} TRACE identified \revision{non-author generated} content with high precision, with only \revision{0.3--6.6\%} of removed text flagged as author\revision{-generated}, and captured 86\% of annotated templated \revision{characters}.
\revision{Compared with models using original notes, models using TRACE-processed notes had cohort-average information extraction F1 differences ranging from $-0.009$ to $+0.004$ and task-specific outcome prediction F1 differences ranging from $-0.011$ to $+0.018$}. In a random sample of 1{,}000 patients at Stanford Health Care, TRACE reduced total chart text by 47.3\% (742.7M characters), corresponding to a mean reduction of 220,167 tokens per patient. Based on 2024 volumes at a large tertiary academic center\revision{ and one query per encounter, projected net savings over three years ranged from \$1.00 million to \$13.58 million across evaluated model pricing schemes, after accounting for initial and annual TRACE pre-processing costs.}

\noindent\textbf{Conclusion:} 
TRACE substantially reduces redundancy in clinical notes while preserving downstream information extraction and prediction performance. These findings demonstrate how underused EHR metadata can be leveraged to reduce LLM inference costs, expand usable longitudinal context, and support more efficient and scalable deployment of clinical AI.
\end{abstract}

%% file: MAIN/010intro.tex

As large language models (LLMs) are increasingly deployed in clinical settings, patient notes have emerged as valuable, rich sources of clinical information. Growing numbers of chart-based assistants allow clinicians to query longitudinal patient records (e.g., Stanford’s ChatEHR, Penn Medicine’s Chart Hero, and Duke’s Scout) \cite{armitage2025chatehr,otto2026charthero,dihi2024scout}, and hospital-wide algorithms have begun to use notes to predict outcomes like sepsis and readmission \cite{pandey2025clinicalt5readmission,shashikumar2025composerllm}. However, naive strategies for utilizing full patient notes can produce prohibitively long sequences within only a few encounters, making it difficult to represent evolving patient states over time. Even as context windows of frontier models grow, processing large volumes of free text remains computationally challenging and can substantially increase cost, compute requirements, and training time. These constraints underscore the need for methods that reduce note length while preserving clinically salient signal.

Clinical notes contain substantial redundant, low-information text introduced by templates, copy/paste, and imported structured data. This phenomenon, commonly referred to as ``note bloat,'' has been extensively documented in prior research. Compounded further by billing incentives and medico-legal documentation pressures, note bloat has led to increasingly duplicative and lengthy notes \cite{hirschtick_copy_paste_2006,weis_copy_2014}. In nearly 3 million outpatient progress notes across 46 specialties, median note length increased 60.1\% and median redundancy increased from 47.9\% to 58.8\% between 2009 and 2018 \cite{rule_length_2021}. In inpatient progress notes, electronic health record (EHR) provenance analyses show that most text is drafted by content-importing tools: in one study of 23{,}630 notes, only 18\% of note text was written by the author, while 46\% was copied, and 36\% was templated \cite{wang_characterizing_2017}. These results align with earlier work, which found that 54\% of the text in progress notes was duplicated from prior notes in the same admission \cite{wrenn_quantifying_2010}. \revision{An analysis of ambulatory physicians' documentation patterns found that decreases in note length were primarily associated with reductions in copied and templated text \cite{apathy_shorter_2023}.} For modern clinical natural language processing (NLP) pipelines, this redundancy can obscure clinically salient information, drive up computational cost, and displace relevant information when inputs are truncated to fit into fixed context windows. 

Health systems have implemented multiple strategies to mitigate redundancy, including institutional guidance on content-importing and user interface features that block out imported or copied content \cite{weis_copy_2014}. In research settings, previous approaches have treated redundancy detection as a text-similarity problem, removing duplicated lines or near-duplicate notes using string-edit distance or set-similarity metrics. For example, Wrenn et al.\ applied a modified Levenshtein edit distance approach to quantify redundancy in sequential notes \cite{wrenn_quantifying_2010}, while Gabriel et al.\ used MinHash-based locality sensitive hashing to remove full notes \cite{gabriel_identifying_2018}.
More recently, Liu et al.\ showed that de-duplication does not degrade performance on prediction tasks and may even improve performance in some settings \cite{liu_note_2022}. However, existing methods typically rely on text similarity comparisons, which are limited to copied segments and may miss imported text from note templates; moreover, these studies do not evaluate downstream performance on LLM-based tasks.

EHR data warehouses contain an underused resource for this problem: note attribution metadata that records the provenance of copied and templated content. Although this information only became readily accessible to clinical users following EHR upgrades in the late 2010s, many hospital databases have logged note-level attribution metadata since at least 2012 \cite{EpicClarityDataDictionary, wang_characterizing_2017, rule_length_2021}. These data identify the source notes for copied text and the templates used to generate templated content. Prior work has used this attribution metadata to \revision{quantify template use, characterize note composition and describe its association with documentation burden\cite{rule_frequent_2022,rotenstein_note_composition_2023,apathy_documentation_dynamics_2023}. These studies focus on documentation practices rather than identifying and removing templated and copied spans within individual notes }for downstream NLP applications. As LLMs increasingly operate over longitudinal records, attribution-guided removal of templated and copied spans offers a practical way to improve data quality, reduce cost, increase computational efficiency, and prevent truncation of longitudinal records exceeding context limits.

To address these gaps, we introduce TRACE (\underline{T}emplate \underline{R}euse \underline{A}nd \underline{C}opied \underline{E}lements), an open-source pre-processing approach that identifies and removes redundant text attributable to templating and copy-pasting behavior, producing condensed notes intended for machine learning applications. We rigorously evaluate TRACE using blinded manual review and comparison to gold standard template annotations. We apply TRACE to four real-world cohorts spanning multiple clinical domains, sites, and tasks, and compare downstream task performance to original unmodified notes. We show that TRACE substantially reduces token count while preserving clinically meaningful content. Moreover, TRACE \revision{preserves} performance on both information extraction and outcome prediction tasks by increasing signal density and reducing truncation of patient medical records at fixed context lengths. We demonstrate the importance of TRACE in real-world clinical settings by showing how TRACE token reductions can substantially decrease projected system-wide costs of note-based LLM deployments. Together, these results support practical adoption of TRACE as a task-agnostic pre-processing tool for clinical notes to increase information density and decrease inference costs \revision{while maintaining downstream task} performance.

%% file: MAIN/040methods.tex
\subsection*{Overview of TRACE Algorithm}
TRACE reduces redundancy in clinical notes using two modules: a Reference Module that uses EHR attribution metadata to identify templated and copied content, and a Frequency Module that removes high-frequency text blocks when attribution data are unavailable. TRACE outputs a condensed note produced by removing templated spans and de-duplicating repeated copied spans. \revision{An illustrative example showing an original note template and the corresponding clinical note with TRACE-identified spans highlighted is provided in Supplementary Figure \ref{sfig:note_example}}.

\xhdr{Reference Module.}
Epic Clarity databases store note-level attribution metadata that link each clinical note to the source texts used in its creation, including note templates and other existing notes from which text was copied and pasted \cite{stanford_starr_ehr, EpicClarityDataDictionary, wang_characterizing_2017, rule_length_2021}. \revision{Template-attribution records associate notes with SmartText and SmartPhrase identifiers (\texttt{note\_smarttext\_ids}, \texttt{note\_smartphrase\_ids}), while copy-tracking records link destination notes to their copied source notes (\texttt{note\_copy\_tracking}). These identifiers allow retrieval of the corresponding template and source-note texts. }These attributions are useful because they substantially narrow the search space for template and copied-text removal by identifying a small set of sources to match against the final note  \revision{(Figure~\ref{fig:method}B)}. However, Clarity does not provide span-level annotations of what information is templated or copied, so span-level provenance must be inferred. Deriving character-level provenance from Clarity alone is non-trivial for several reasons: template insertion may be partial, multiple sources may contribute overlapping content, and templated or copied text may be minimally edited after insertion, obscuring where sourced text begins and ends in the final note.

To translate note-level attributions into span-level matches, TRACE aligns each attributed source (template text and any copied source notes) to the final note using the Ratcliff–Obershelp algorithm. For each source–note pair, the algorithm identifies matching text blocks by recursively finding the longest common substring and then repeating the procedure on the remaining unmatched regions, yielding a set of candidate templated or copied spans. Matching is performed independently for each attributed source (that is, for a note with 
$k$ attributed sources, alignment is run 
$k$ times), and the resulting spans are pooled (Figure~\ref{fig:method}A). TRACE then merges spans separated only by whitespace and discards spans shorter than a user-defined threshold. We selected a threshold of 50 characters to avoid removing short contextual fragments and to target longer text chunks that are not specific to the patient encounter \revision{(Supplementary Note~\SIref{3})}. Finally, TRACE removes all templated spans and de-duplicates copied spans, retaining the first instance.

\xhdr{Frequency Module.}
To address incomplete or unavailable attributions (e.g., notes drafted outside the Epic interface, notes written before attribution metadata were recorded, or datasets without provenance metadata such as MIMIC-III), TRACE applies reference-free frequency-based de-duplication to all notes after the Reference Module. Building on prior work, we focused on frequency-based de-duplication to remove large spans of text \cite{li_memorization_2025}. The Frequency Module chunks notes into both sentences, delimited by a period, and paragraphs, delimited by new lines. Chunks were normalized by removing multiple spaces and converting to lowercase, then filtered to spans with at least 50 non-space characters. Note chunks were labeled as templated if found in more than 5 patients and labeled as copied if found more than once in a single patient. \revision{Templated spans were removed, while copied spans were de-duplicated, retaining the first occurrence in each patient. Semi-structured elements, such as Physical Exam macros with editable checkboxes, were not explicitly detected by either the Reference or Frequency modules. This text can be detected by the Reference Module when these elements are included in note templates and matching spans meet the length threshold, or by the Frequency Module when normalized chunks meet the repetition and length thresholds. }The overall procedure is detailed in Figure \ref{fig:method}A. A runtime analysis of TRACE can be found in \revision{Supplementary Note~\SIref{1}}. 

\subsection*{Evaluation of Template Identification}
We evaluated TRACE’s ability to identify templated text using two complementary character-level assessments: (1) a blinded manual review of template spans identified by TRACE and (2) a comparison against human-annotated template annotations derived from the Epic Clarity attribution metadata. For these evaluations, we randomly sampled 10\revision{0} clinical notes at Stanford Health Care (SHC) and Lucile Packard Children's Hospital (LPCH)\revision{, SHC's affiliated children's and obstetric hospital.} 

\xhdr{Blinded Character Analysis.}
To estimate the precision of each TRACE module and characterize whether removed text was templated or author-entered, we conducted a blinded manual character-level review. \revision{Two} board-certified physician\revision{s} reviewed 150 spans from each of three sets: spans identified by the Reference Module, spans identified by the Frequency Module, and randomly generated spans matched in length to TRACE spans. The annotator\revision{s were} blinded to each span's source to prevent bias. For each set, we report the fraction of total characters that \revision{each annotator} classified as ``true'' templated text, author-generated text, imported structured data (e.g. labs, vital signs), or indeterminate. \revision{We additionally report inter-annotator agreement using Cohen's $\kappa$.}

\xhdr{Attribution Metadata Template Detection.}
To estimate the recall of templated text, we created gold-standard span annotations by manual review of notes and their associated templates. \revision{Annotators} were provided each note together with the templates linked to that note in Epic Clarity. Notably, these are the same inputs to identify templated text used by the Reference Module. These templates included both the original links to structured data elements and predefined templated text blocks. Annotators labeled all characters in the note as (i) author-entered, (ii) templated text, or (iii) imported structured data (e.g., refreshable links that automatically import age, vitals, or recent laboratory values). Candidate template and structured-data spans were first proposed by \revision{GPT-4.1 (Supplementary Figure \ref{sfig:prompt_spans}), selected because it reliably followed the annotation instructions and produced consistently formatted output during development of the annotation workflow. Proposals were then refined by two non-clinical study authors (J.L.C. and A.A.).} A board-certified physician (C.S.) \revision{reviewed these draft annotations and }provided the final character-level annotations. 

We evaluated the TRACE Reference Module using these gold standard template annotations. Recall was calculated for spans at least 50 characters long, consistent with TRACE’s minimum span length and to focus evaluation on longer, encounter-nonspecific text while preserving short contextual elements (e.g., physical exam or subsection headers). Precision was calculated across all text spans identified by human annotators, since a single, long TRACE span can cover multiple short spans identified in the gold-standard dataset. We do not report F1 due to the differences in the minimum span lengths for precision and recall calculation. \revision{To evaluate performance variations with alternative minimum span lengths, we additionally evaluated thresholds of 25, 50, and 75 characters. The same reference annotations and eligible reference spans were used at all three thresholds, ensuring that recall was evaluated against a fixed reference set. We assessed text removal in 1,000 randomly sampled patient timelines, template-detection precision and recall in the annotated 100-note subset, and downstream embedding-based PPH prediction (Supplementary Note~\SIref{3}).}  
Because the gold-standard annotations were derived from Epic Clarity attributions, they capture only attribution-linked templated content; the Frequency Module is designed to remove redundancy not represented in Clarity, so precision and recall for this module could not be meaningfully computed against this reference.

Furthermore, we did not evaluate copied-text identification in either the blinded span review or the attribution-based evaluation, as there is no reliable reference for copied content. Manual adjudication was not feasible because copied content can be as small as a single word or character, and comparable overlap between source and destination notes can occur by chance. Furthermore, because copied text is de-duplicated rather than fully deleted, occasional false positives are less consequential than for templated text, where flagged content is removed entirely.

\subsection*{Evaluation Datasets \& Downstream Tasks}

\revision{We evaluated the impact of TRACE on information extraction and outcome prediction using four datasets from two health systems, selected to span distinct documentation patterns (Figure~\ref{fig:validation}B). Liver transplant recipients at SHC have long, complex longitudinal records; obstetric patients at SHC and LPCH combine outpatient prenatal documentation with a concentrated delivery admission; and inpatient admissions from SHC and the publicly available MIMIC-III v1.4 dataset from Beth Israel Deaconess Medical Center (BIDMC) \cite{johnson_mimic-iii_2016,PhysioNet-mimiciii-1.4,goldberger_physiobank_2000} enable cross-institutional evaluation of 30-day readmission prediction.} We additionally characterized TRACE in a task-agnostic sample of 1{,}000 patients randomly selected from SHC and LPCH Labor \& Delivery admissions.

Attribution metadata availability differed across datasets. SHC and LPCH notes were sourced from Epic Clarity and included the note-level template and copy-tracking metadata used by the Reference Module. These metadata were unavailable in MIMIC-III, where TRACE used only the Frequency Module. Cohort construction and downstream tasks are described below.

\xhdr{Liver Transplant.}
We studied a cohort of \revision{279} liver transplant recipients \revision{with a history of hepatocellular carcinoma }transplanted between 2008 and 2018, sampled according to the protocol described in \cite{adeniji_impact_2020}. Clinical variables relevant for evaluating transplant candidacy were extracted using clinician chart review \cite{adeniji_impact_2020}. These labels include the presence of 1\revision{3} comorbidities and 5-year overall survival (OS). OS was defined as time from transplant to death from any cause. Information extraction (IE) was performed on all notes up to one year after the patient's \revision{hepatocellular carcinoma} diagnosis date, and OS prediction was performed on all notes through the transplant date. Patients without notes in the 1-year post-diagnosis window were excluded from the IE experiments. 

\xhdr{Obstetrics.}
The obstetric cohort included all deliveries at SHC and LPCH from 2016-2025, totaling 42,130 deliveries \revision{(34,360 unique patients)}. For information extraction, notes were extracted 1 year before delivery through 1 week after delivery. For 100 deliveries, gold standard attributes were extracted via clinician chart review performed by licensed physicians (a Maternal-Fetal Medicine fellow and an Obstetrics attending). These are clinically relevant factors for this cohort that can be challenging to extract reliably from structured data. They included 2 obstetric history factors and 5 intrapartum/delivery characteristics. \revision{For outcome prediction, we defined postpartum hemorrhage (PPH) as cumulative quantified blood loss (QBL) $\geq$1{,}000~mL within 24 hours after delivery, consistent with the American College of Obstetricians and Gynecologists threshold \cite{acog_pph_2017}. When QBL was missing or recorded as zero, blood transfusion between delivery and hospital discharge was used as a surrogate marker.} For PPH prediction, notes from \revision{12} months before delivery through 1 hour before delivery were used as input.  

\xhdr{Inpatient.}
We constructed datasets of inpatient admissions from two healthcare systems (SHC, BIDMC) to evaluate 30-day readmission, a quality metric for hospital-based care \cite{ReducingReadmissions}. For SHC, we randomly sampled 50{,}000 inpatient encounters from discharge summaries written between 2018 and 2025 (38{,}928 unique patients), excluding summaries that mentioned death using regular-expression matching (Supplementary Note\revision{~\SIref{2}}). For each sampled admission, all clinical notes associated with that admission were extracted. A 30-day readmission was defined as the occurrence of at least one inpatient admission within 30 days after discharge; \revision{admissions to outside health systems were not captured.} We also evaluated TRACE on admissions from MIMIC-III v1.4 \cite{johnson_mimic-iii_2016,PhysioNet-mimiciii-1.4,goldberger_physiobank_2000}, excluding newborn and death encounters. In total, there were 45,135 admissions, representing 34,560 unique patients. 30-day readmission was defined as the occurrence of at least one non-elective inpatient admission within 30 days following discharge from the index admission, using the protocol described in \cite{clinicalbert}. We refer to the two inpatient cohorts as Inpatient-SHC and MIMIC-III.

\xhdr{\revision{Task}-Agnostic SHC Cohort.}
In addition to these task-specific datasets, we randomly sampled 1{,}000 patients from SHC and from Labor \& Delivery admissions at LPCH\revision{ and applied TRACE to each patient's complete longitudinal note history (435{,}797 notes) to estimate task-agnostic text and token reduction. The 100 notes used for template-identification evaluation and blinded span review were drawn from this sample.}

\subsection*{Downstream Evaluation Framework}
TRACE downstream performance was assessed on information extraction (IE) and clinical outcome prediction.\revision{ IE involved recovering clinical variable values explicitly documented in the available notes, whereas outcome prediction used notes available before a defined prediction time to predict a subsequent clinical outcome. We evaluated IE using zero-shot LLM inference and outcome prediction using either zero-shot LLM inference or classifiers trained on note embeddings.}

\xhdr{Zero-shot Information Extraction.}
\revision{For the liver transplant and obstetric cohorts, we used cohort-specific prompts to extract target clinical variables from each patient’s note history. }Each clinical note was formatted to contain a note identifier, note date, and note text. Notes for a given patient were ordered in ascending time and batched to fit in \revision{the 1M-token context window. When multiple batches were required, the previously extracted values were supplied alongside the next batch of notes, allowing the model to update its answers using the additional information (Supplementary Figures \ref{sfig:prompt_ie} and  \ref{sfig:prompt_ie_update}). The final extracted values were compared with clinician-annotated reference labels. }

\xhdr{Zero-shot Outcome Prediction.}
Zero-shot \revision{LLM }inference was conducted \revision{to predict five-year overall survival in the liver transplant cohort using clinical notes available through the transplant date. The target was subsequent survival status rather than information already documented in the input notes. Notes were first processed and ordered chronologically as described above}. To handle the large number of patient notes, inference was performed on the most recent tokens per patient that fit within the 1M-token context window (Supplementary Figure \revision{\ref{sfig:prompt_zero}). Zero-shot outcome prediction was only evaluated in the liver transplant cohort because LLM inference across all encounters in the larger cohorts would have been prohibitively expensive. All zero-shot IE and outcome-prediction inference used Gemini-2.5-Pro, which was selected for having the largest context window among the frontier models available to our team at the time of analysis (1 million tokens).}

\xhdr{Embedding-based Outcome Prediction.}
For cohorts with greater than 1{,}000 patients (Obstetrics, Inpatient-SHC, and MIMIC-III), we trained \revision{gradient boosted classifiers on note embeddings to predict postpartum hemorrhage or 30-day readmission} \cite{liu_note_2022,clinicalbert, amrollahi2021contextual}. \revision{This approach enabled supervised prediction across the larger cohorts without requiring generative LLM inference for every encounter and allowed us to assess TRACE in a complementary modeling framework. }Patient-level representations were constructed by embedding the 32,000 most recent \revision{note} tokens\revision{ prior to the cohort-specific prediction cutoff} with Qwen3-Embedding-8B \cite{qwen3embedding}. For each cohort, \revision{patients were assigned to training and test sets with an outcome-}balanced 85-15 train-test split, and hyperparameters were tuned with cross-validation on the training set. \revision{Classification thresholds were selected to maximize F1 using a validation subset comprising }20\% of the training split. 

\xhdr{Statistical Analysis.}
\revision{We estimated paired differences in downstream performance, defined as the metric calculated using TRACE-processed inputs minus the same metric calculated using Original inputs. For IE, we calculated differences in F1 for each extraction task. We also calculated mean differences separately for the liver transplant and obstetric cohorts, assigning equal weight to each of the 13 and 7 tasks, respectively. For embedding-based prediction of postpartum hemorrhage and 30-day readmission in SHC and MIMIC-III, we calculated differences in AUROC and F1. For zero-shot prediction of five-year overall survival, we calculated differences in F1. Changes in sensitivity and specificity were also assessed using the logistic regression models described below.}

\revision{We estimated 95\% confidence intervals using 10{,}000 paired bootstrap replicates within each cohort, sampling patients with replacement except for obstetric tasks, for which deliveries were sampled with replacement. We recalculated each metric separately for Original and TRACE on the resampled observations, then subtracted the Original estimate from the TRACE estimate. For cohort-average IE estimates, we calculated this difference for each extraction task and averaged the differences in F1 scores across the cohort's tasks. Confidence limits were the 2.5th and 97.5th percentiles of the resulting bootstrap distributions. Fitted classifiers and classification thresholds were held fixed during resampling of the test predictions. These confidence intervals were not adjusted for multiple comparisons and provide individual rather than simultaneous coverage across endpoints.}

\revision{We conducted nine post-hoc non-inferiority tests using an absolute margin of 0.05. The absolute margin of 0.05 was selected post-hoc by the investigators as a pragmatic tolerance for performance loss when evaluating the benefits of text reduction. The hypotheses were grouped into three families: cohort-average IE F1 for liver transplant and obstetrics (two hypotheses); AUROC for embedding-based prediction of postpartum hemorrhage and readmission in SHC and MIMIC-III (three hypotheses); and F1 for these three outcome-prediction tasks and zero-shot prediction of five-year overall survival (four hypotheses).  These families addressed extraction performance, discrimination, and classification performance at the selected decision thresholds, respectively. For IE, the hypotheses concerned mean F1 differences across extraction tasks within each cohort. }

\revision{For each endpoint, we tested $H_0:\Delta \leq -0.05$ against $H_1:\Delta > -0.05$, where $\Delta$ denotes TRACE minus Original performance. The null hypothesis therefore represented a performance loss of at least 0.05. We calculated one-sided Wald p-values using: }
\begin{equation}
\label{eq}
Z = \frac{\widehat{\Delta}+0.05}
{\widehat{\mathrm{SE}}_{\mathrm{boot}}},
\end{equation}
\revision{where $\widehat{\Delta}$ was the observed paired performance difference and
$\widehat{\mathrm{SE}}_{\mathrm{boot}}$ was the standard deviation of the paired differences across the bootstrap replicates described above. P-values were obtained from the upper tail of the standard normal distribution. We applied Holm's procedure separately within each hypothesis family and concluded non-inferiority when the adjusted one-sided p-value was below 0.025. This procedure controlled family-wise error.}

\revision{We fit exploratory logistic regression models to assess associations of TRACE-attributable text reduction with sensitivity and specificity of downstream tasks. Models were fit separately among reference-positive observations, for which a correct prediction contributes to sensitivity, and reference-negative observations, for which a correct prediction contributes to specificity. }Models were estimated by maximum likelihood with patient-level cluster-robust (sandwich) standard errors to account for repeated observations within patients. \revision{We fit the following model separately among reference-positive and reference-negative observations:}
\begin{equation}
\label{eq:arm}
\text{logit}\!\left(\Pr(\text{correct}_i = 1)\right)
= \beta_0
+ \beta_1 \,\text{arm}_i
+ \beta_2 \,\text{feature\_prevalence}_i,
\end{equation}
where $i$ indexes a unique prediction, $\text{correct}_i$ indicates whether the prediction \revision{matched the reference label}, $\text{arm}_i$ is an indicator for TRACE-processed input (vs.\ Original), and $\text{feature\_prevalence}_i$ denotes the prevalence of the target label within the sample. \revision{Feature prevalence was included only in IE models; the prevalence term was omitted from outcome-prediction models. The same stratification on positive and negative reference labels, as well as the same covariate specifications were used in the text-removal models below. Holm correction was applied to the TRACE coefficient within each task group.}

\revision{Among TRACE-processed inputs, we also assessed whether the percentage of text removed was associated with sensitivity and specificity, adjusting for original input length and, for IE, feature prevalence:}\begin{equation}
\label{eq:length}
\begin{aligned}
\operatorname{logit}\!\left[\Pr\!\left(\mathrm{correct}_j = 1\right)\right]
={}& \alpha_0
+ f\!\left(\mathrm{delta\_length}_j\right) \\
&+ g\!\left(\mathrm{total\_length}_j\right)
+ \alpha_1 \mathrm{feature\_prevalence}_j
\end{aligned}
\end{equation}
\revision{where $j$ indexes a TRACE prediction, $\mathrm{delta\_length}_j$ denotes the percentage of text removed
by TRACE, and $\mathrm{total\_length}_j$ denotes the standardized
original input length.  The functions $f$ and $g$ were restricted cubic splines with five knots each: three interior knots at the 25th, 50th, and 75th percentiles and two boundary knots at the minimum and maximum of each variable’s distribution, allowing nonlinear associations with prediction accuracy. For the liver transplant survival-sensitivity analysis, the spline model could not be estimated, so we used a logistic regression model with linear terms, requiring fewer parameters.}

\revision{We assessed the overall text-removal association by jointly testing the coefficients of the text-removal terms using a Wald test, with Holm correction across cohorts within each task group. To describe the fitted associations with sensitivity and specificity, we calculated adjusted predicted probabilities of a correct prediction across removal percentages, separately for reference-positive and reference-negative observations. Original input length was held at its mean. For IE analyses, feature prevalence was held at the task-specific prevalence within each cohort. Pointwise 95\% confidence intervals were calculated using the delta method and the patient-level cluster-robust covariance matrix. These analyses characterize associations rather than causal effects of text removal.}

\revision{As a complementary descriptive analysis that does not rely on a regression model, we stratified cases (patients, or deliveries for obstetric tasks) within each cohort into quartiles of the percentage of text removed by TRACE, from the least (Q1) to the most (Q4) text removed, and estimated the paired F1 difference (TRACE minus Original) within each quartile. Quartile boundaries were defined within each cohort and therefore correspond to different ranges of text removal across cohorts. Within each quartile, we drew 1{,}000 bootstrap replicates, resampling cases with replacement while keeping Original and TRACE predictions paired, and recalculated F1 for each replicate. For IE, feature-level F1 differences were averaged across extraction tasks within each replicate. Predictions and classification thresholds were held fixed. We report the observed difference with 95\% percentile confidence intervals (2.5th and 97.5th percentiles of the bootstrap distribution).}

\revision{The originally planned analyses included metadata-assisted annotation of 100 notes, paired downstream comparisons of Original and TRACE inputs for the evaluation cohorts, and secondary exploratory regression analyses. Cohort sizes reflected available eligible patients, while the annotation target reflected available annotation resources. Blinded span review was added during the study. As a post-hoc analysis, we added minimum-span sensitivity analyses, incorporated spline terms into the regression models, and conducted exploratory post-hoc non-inferiority tests using a margin selected after initial performance results were available.}

\xhdr{Cost and utilization modeling.}
We estimated the economic impact of TRACE on large-scale LLM deployment by projecting inference cost savings associated with reduced input token volume. Using the random 1{,}000-\revision{patient} sample from SHC, we applied TRACE and computed the mean per-patient token reduction using the Qwen3-8B tokenizer. We then converted token savings to dollars using \revision{provider API list prices as of September 21, 2026: GPT-6 Astra (\$10 per million tokens), GPT-5.6 Terra (\$2 per million tokens) \cite{openai_api_pricing}, Claude Opus 5 (\$5 per million tokens) \cite{anthropic_api_pricing}, and Gemini 3.8 Flash (\$0.75 per million tokens) \cite{google_gemini_api_pricing}.} While deployed LLM tools may summarize the patient chart and run inference over generated summaries in practice, we make the assumption that all queries make a single pass over the patient's \revision{longitudinal record. Annual gross savings were calculated using 2{,}058{,}497 FY2024 Stanford Health Care encounters} derived from published utilization data \cite{fitch_stanford_2025}, \revision{and costs calculated assuming either one query every four encounters, one query per encounter, or five queries per encounter.}

\revision{Pre-processing costs were estimated from the sample's aggregate CPU-hours and estimated memory-hours using Google Cloud E2 1-year committed-use pricing, accessed September 21, 2026 \cite{google_cloud_compute_pricing}, assuming an existing, sufficiently utilized institutional agreement. We extrapolated the sample's note volume and processing costs to approximately 3 million patients in SHC Clarity \cite{stanford_starr_ehr}, assuming that the sample represented this population in average timeline length, token reduction, and processing cost. Annual new-note volume was assumed to equal a fixed, non-compounding 9\% of baseline volume based on the average number of annual new notes at SHC 2022-2024. Processing costs were assumed to scale with note count, with new notes processed using retained historical information without rerunning TRACE on the full corpus. Cumulative net savings over one or three years were calculated by subtracting the initial pre-processing cost and one or three annual processing costs from cumulative gross savings. Annual query volume, mean token reduction, and prices were held constant, without prompt-caching or batch inference discounts. Storage and other operational costs were excluded. For more information on the assumptions used for these calculations, see Supplementary Figure \ref{sfig:cost}.}

\subsection*{Ethics approval.}
This retrospective study was approved by the Stanford University Institutional Review Board (IRB-78065 and IRB-50290) with a waiver of informed consent and HIPAA authorization. Data from SHC and LPCH were accessed and analyzed under these protocols. Use of MIMIC-III was conducted under the PhysioNet credentialed access and data use agreement \cite{johnson_mimic-iii_2016,PhysioNet-mimiciii-1.4,goldberger_physiobank_2000}.

\subsection*{Data and code availability.}
Documentation to run TRACE is available at \url{https://github.com/alsentzerlab/TRACE}. Data for the three Stanford Health Care and Lucile Packard Children's Hospital cohorts (Liver Transplant, Obstetrics, Inpatient-SHC) are unable to be released at this time because they contain protected health information. MIMIC-III v1.4 \cite{johnson_mimic-iii_2016,PhysioNet-mimiciii-1.4,goldberger_physiobank_2000} is available at \url{https://physionet.org/content/mimiciii/1.4/}

%% file: MAIN/020results.tex
\subsection*{TRACE accurately identifies templated text in clinical notes}
\label{res:eval}
We conducted a blinded evaluation of template spans identified by TRACE to evaluate the precision of template text removal. \revision{Two annotators reviewed TRACE-selected and randomly sampled spans, classifying each character as belonging to templated text, structured data, manually authored text, or an uncertain category. Overall character-level agreement was 78.88\% (Cohen’s $\kappa = 0.63$). Although absolute estimates differed between annotators, the pattern across groups was consistent. Ranges reported below represent the two annotators’ point estimates; annotator-specific estimates and 95\% confidence intervals are provided in Supplementary Table \ref{tab:gold_span_bootstrap}. Templated content comprised the majority of characters in spans identified by both modules (Reference Module 77--91\%; Frequency Module 80--90\%). Only 0.3--6.6\% of characters identified by each module were flagged as author-entered and 0.1--3.1\% were automatically imported structured data (Reference Module 0.1--2.3\%; Frequency Module 1.8--3.1\%). The remainder of characters comprising spans flagged by TRACE were of uncertain origin, in which the physician annotators were unable to determine whether they were author-entered or templated (Reference Module 7--14\%; Frequency Module 7--16\%). The heavily templated composition of spans from both TRACE modules contrasts with that of the random spans, in which only 19--31\% of characters were identified as templated and a higher proportion were author-written (15--24\%) and structured data (37--42\%)}.

We further evaluated the Reference Module by comparing flagged spans to human-generated gold-standard template annotations. The Reference Module achieved a character-level precision of 0.96 (95\% CI: 0.93-0.98) and recall of 0.84 (95\% CI: 0.75-0.90) (Supplementary Table \ref{tab:ablation_accuracy}). Interestingly, incorporating both modules in the full TRACE pipeline leads to a slight increase in recall, for a total of 0.86 (95\% CI: 0.78-0.92) (Supplementary Table \ref{tab:ablation_accuracy}), even though the Frequency Module was not explicitly designed to identify templates present in Epic Clarity metadata attributions. 

\revision{To examine the sensitivity of text removal to the minimum span threshold, we conducted an ablation study where the minimum span threshold was varied. Increasing the threshold from 25 to 50 to 75 characters reduced corpus-level text removal from 49.0\% to 47.3\% to 45.1\% (Supplementary Table \ref{tab:ablation_removal}). Reference Module recall decreased from 0.86 to 0.84 to 0.77, with precision of approximately 0.96 at each threshold, and full-pipeline recall decreased from 0.88 to 0.86 to 0.79, with precision of 0.88, 0.88, and 0.89 (Supplementary Table \ref{tab:ablation_accuracy}). For PPH prediction, TRACE AUROC was 0.748, 0.751, and 0.745 at thresholds of 25, 50, and 75 characters, respectively, compared with 0.747 using Original notes, and F1 was 0.407, 0.416, and 0.399, compared with 0.398 (Supplementary Table \ref{tab:ablation_pph}).}

\subsection*{TRACE enables analysis of templated and copied text in real-world cohorts}
\label{res:random}
To assess the prevalence of templated and copied text spans in a specialty-agnostic cohort, we applied TRACE to the full longitudinal note histories of 1{,}000 randomly sampled patients from SHC and the Labor and Delivery adult admissions at LPCH. In total, there were \revision{435{,}797 } notes written from 1998-2025. TRACE flagged at least one span in \revision{263{,}850} notes, with \revision{209{,}821} flagged by the Reference Module and \revision{219{,}797} by the Frequency Module. Across all notes in the sample, TRACE removed 47.3\% of characters (corpus level), and the mean reduction per patient timeline was 38.9\% (Figure~\ref{fig:random}A; \revision{Supplementary Table~\ref{tab:ablation_removal}}). 

Among characters flagged by TRACE, 77\% were labeled as copied and an additional 3\% were labeled as both copied and templated text (Figure \ref{fig:random}B). \revision{Approximately 45\% of characters flagged by TRACE were identified by the Reference Module alone, 22\% by the Frequency Module alone, and 33\% by both modules (Figure \ref{fig:random}C)}. This suggests that EHR attribution metadata is a valuable resource for identifying templated and copied text, particularly when a template is underrepresented or when imported text is edited. However, the attribution metadata may be incomplete for notes that were copied from other institutions, drafted outside of Epic, or published prior to when note attributions were recorded in the EHR. \revision{The Frequency Module identified approximately one fifth of flagged characters that were not identified by the Reference Module, demonstrating its complementary contribution.}

We also examined the prevalence of templated and copied text by note type. Out of the ten most frequent note types \revision{in the 1{,}000 patient sample, Patient Instructions, Discharge Instructions, Procedures, and Telephone Encounters contained the highest proportions of templated text (62\%, 59\%, 27\%, and 27\%, respectively), whereas Progress Notes, Anesthesia Preprocedure Evaluations, Discharge Summaries, and H\&Ps contained the highest proportions of copied text (37\%, 27\%, 22\%, and 22\% respectively).} These results are consistent with previous findings that note \revision{composition and redundancy vary across specialties \cite{rotenstein_note_composition_2023, rule_length_2021}}. \revision{In Figure \ref{fig:random}D, we show the distribution of the per-note percentage of characters flagged as templated, copied, or either (total) for the most common note types in the 1{,}000-patient sample (Supplementary Tables \ref{tab:notetype_total}--\ref{tab:notetype_copyforward})}.

\subsection*{\revision{TRACE decreases note length while maintaining information extraction performance.}}
TRACE reduced note length across cohorts by 2\revision{4}\% (Liver Transplant), 31\% (Obstetrics), 30\% (Inpatient-SHC), and 12\% in MIMIC-III (Frequency Module only) (Figure~\ref{fig:ie}B). We evaluated information extraction on the subset of the Liver Transplant (N=276) and Obstetrics (N=100) cohorts with gold-standard clinician annotations. 
While performance for individual tasks varied, \revision{mean F1 differences (TRACE minus Original) were 0.0040 (95\% CI $-0.0047$ to 0.0126) in the Liver Transplant cohort and $-0.0086$ (95\% CI $-0.0229$ to 0.0043) in the Obstetrics cohort. Both comparisons met the non-inferiority criterion (Holm-adjusted $p<0.001$; Figure~\ref{fig:ie}A; Supplementary Tables~\ref{tab:trace-ie-f1-deltas} and~\ref{tab:trace-noninferiority}). Stratified by within-cohort quartile of text removed, average IE F1 differences were small in each quartile (Figure~\ref{fig:ie}C). In the Liver Transplant cohort, differences ranged from $-0.0105$ (95\% CI $-0.0327$ to 0.0105) in the lowest-removal quartile to 0.0131 (95\% CI $-0.0001$ to 0.0279) in the third quartile, and were 0.0113 (95\% CI $-0.0074$ to 0.0296) in the highest-removal quartile. In the Obstetrics cohort, the lowest- and highest-removal quartiles both showed small decreases ($-0.0229$, 95\% CI $-0.0571$ to 0; and $-0.0171$, 95\% CI $-0.0514$ to 0), whereas the middle quartiles showed differences near zero.} Finally, information extraction on TRACE-processed notes outperformed a length-matched baseline that removed random spans (Supplementary Figure \ref{sfig:random}).

\revision{We used two complementary regression models to examine associations between TRACE processing and IE accuracy. First, we compared predictions made using TRACE-processed versus Original notes, adjusting for feature prevalence (Equation \ref{eq:arm}). TRACE processing was not significantly associated with sensitivity in either the Liver Transplant cohort (adjusted OR 1.05, 95\% CI 0.97--1.14; Holm-adjusted $p=0.42$) or the Obstetrics cohort (adjusted OR 0.93, 95\% CI 0.83--1.04; Holm-adjusted $p=0.42$). Similarly, TRACE processing was not significantly associated with specificity in either the Liver Transplant cohort (adjusted OR 1.00, 95\% CI 0.95--1.06; Holm-adjusted $p=0.87$) or the Obstetrics cohort (adjusted OR 0.91, 95\% CI 0.77--1.08; Holm-adjusted $p=0.57$) (Supplementary Tables \ref{tab:trace-logistic-arm-specificity} and \ref{tab:trace-logistic-arm-sensitivity}). Second, among predictions made using TRACE-processed notes, we examined whether the percentage of text removed was associated with accuracy, adjusting for original input length and feature prevalence. Restricted cubic splines allowed nonlinear associations with text removal and original input length (Equation \ref{eq:length}). For sensitivity, the joint test of the text-removal spline was significant before, but not after, Holm correction in the Obstetrics cohort (unadjusted $p=0.049$; Holm-adjusted $p=0.098$), whereas the corresponding test was not significant in the Liver Transplant cohort (Holm-adjusted $p=0.39$). For specificity, neither cohort showed a significant association (Holm-adjusted $p=1.0$ for both cohorts; Supplementary Tables~\ref{tab:trace-spline-tests-sensitivity} and~\ref{tab:trace-spline-tests-specificity}; Supplementary Figure~\ref{sfig:spline-ie}).}

\subsection*{\revision{TRACE produces similar clinical outcome prediction performance estimates}}
We evaluated the impact of TRACE-processing on predictive performance using two evaluation paradigms: (1) embedding-based classifiers trained on note representations for cohorts with sufficient training data and (2) zero-shot inference with a language model.

\revision{For embedding-based classification, models trained on TRACE-processed notes achieved AUROC scores of 0.751, 0.803, and 0.708 for PPH, 30-day readmission (Inpatient-SHC), and 30-day readmission (MIMIC-III), respectively, compared with 0.747, 0.802, and 0.712 using Original notes (Figure \ref{fig:outcome}A; Supplementary Table \ref{tab:out_bootstrap_auroc}). Paired TRACE-minus-Original AUROC differences were 0.0047 (95\% CI $-0.0034$ to 0.0127), 0.0003 (95\% CI $-0.0052$ to 0.0059), and $-0.0033$ (95\% CI $-0.0147$ to 0.0080), respectively (Supplementary Table \ref{tab:trace-outcome-deltas}). All three comparisons met the non-inferiority criterion (Holm-adjusted $p<0.001$; Supplementary Table~\ref{tab:trace-noninferiority})}.

Consistent with AUROC, classifiers trained on embeddings of TRACE notes achieved F1 scores of \revision{0.42}, 0.57, and 0.23 for PPH, 30-day readmission (Inpatient-SHC), and 30-day readmission (MIMIC-III), respectively, compared with \revision{0.40}, 0.57, and 0.22 using Original notes (Figure \ref{fig:outcome}B, Supplementary Table \ref{tab:out_bootstrap_f}). Models trained on TRACE-processed notes achieved AUPRCs of \revision{0.38}, 0.60, and 0.16 for PPH, Inpatient-SHC readmission, and MIMIC-III readmission, respectively, compared with \revision{0.38}, 0.60, and 0.16 using Original notes. In both arms, AUPRC exceeded twice the outcome prevalence for each task (Supplementary Table \ref{tab:outcome_auprc}). \revision{Logistic regression models (Equation~\ref{eq:arm}) indicated small shifts in the balance between sensitivity and specificity at the selected thresholds. TRACE was associated with higher sensitivity and lower specificity for 30-day readmission in Inpatient-SHC (sensitivity OR 1.25, 95\% CI 1.16--1.34; specificity OR 0.77, 95\% CI 0.74--0.81) and MIMIC-III (sensitivity OR 1.24, 95\% CI 1.07--1.43; specificity OR 0.86, 95\% CI 0.81--0.91), and with lower sensitivity and higher specificity for PPH (sensitivity OR 0.78, 95\% CI 0.71--0.86; specificity OR 1.47, 95\% CI 1.40--1.55) (all Holm-adjusted $p<0.01$; Supplementary Tables~\ref{tab:trace-logistic-arm-sensitivity} and \ref{tab:trace-logistic-arm-specificity})}.

In the zero-shot setting, prediction of 5-year OS in the Liver Transplant cohort using TRACE-processed notes achieved an F1 score of \revision{0.35, compared with 0.36 using Original notes ($\Delta$F1 $-0.011$, 95\% CI $-0.032$ to 0.007; Figure~\ref{fig:outcome}B; Supplementary Tables~\ref{tab:trace-outcome-deltas} and~\ref{tab:out_bootstrap_p}--\ref{tab:out_bootstrap_f}). TRACE was not significantly associated with sensitivity (OR 0.58, 95\% CI 0.27--1.24) or specificity (OR 1.02, 95\% CI 0.84--1.25) (Supplementary Tables~\ref{tab:trace-logistic-arm-sensitivity} and \ref{tab:trace-logistic-arm-specificity}). All seven outcome-prediction endpoints met the non-inferiority criterion (Supplementary Table~\ref{tab:trace-noninferiority}). Together, these results indicate that TRACE maintained discrimination and F1 across outcome prediction tasks, with task-specific shifts in the balance between sensitivity and specificity at the selected thresholds.}

\revision{We also examined whether the amount of text removal by TRACE affects downstream outcome prediction performance. For each cohort, we fitted logistic regression models among predictions made using TRACE-processed notes, adjusting for original input length. Restricted cubic splines allowed nonlinear associations with text removal and original input length (Equation \ref{eq:length}). For Overall Survival sensitivity, we used a logistic regression model with linear terms because the spline model was singular. Text removal was not significantly associated with sensitivity or specificity for OS (Holm-adjusted $p=0.07$ and $p=0.64$, respectively) or PPH (Holm-adjusted $p=0.33$ and $p=0.082$, respectively). For 30-day readmission, text removal was significantly associated with both sensitivity and specificity in Inpatient-SHC (Holm-adjusted $p=1.2\times10^{-4}$ and $p=2.2\times10^{-25}$, respectively) and MIMIC-III (Holm-adjusted $p=0.013$ and $p=1.5\times10^{-26}$, respectively) (Supplementary Tables \ref{tab:trace-spline-tests-sensitivity} and \ref{tab:trace-spline-tests-specificity}; Supplementary Figure~\ref{sfig:spline-pred}). Across analyses, the relationship between text removal and performance varies across cohorts and endpoints. Stratified by within-cohort quartile of text removed, F1 differences across the four outcome-prediction tasks ranged from $-0.0227$ to 0.0296, and 95\% confidence intervals included zero in 15 of 16 quartiles (Figure~\ref{fig:outcome}C). The exception was the third quartile for PPH (0.0296, 95\% CI 0.0002 to 0.0570), which favored TRACE. Differences did not decline consistently with increasing text removal. In the highest-removal quartile, differences were 0.0038 (95\% CI $-0.0095$ to 0.0182) for survival, 0.0222 ($-0.0054$ to 0.0479) for PPH, $-0.0195$ ($-0.0389$ to 0.0002) for Inpatient-SHC readmission, and 0.0120 ($-0.0232$ to 0.0467) for MIMIC-III readmission. Survival estimates, based on 69--70 patients per quartile, were the least precise. Thus, although text removal was associated with sensitivity and specificity for 30-day readmission, the direction of these associations differed across cohorts and endpoints and did not correspond to consistently lower F1 among patients with more text removed.}

\subsection*{TRACE token reductions can substantially reduce spending on LLM inference}
Across the 1{,}000-patient task-agnostic sample, TRACE reduced total chart text from 1{,}570{,}831{,}445 characters to 828{,}132{,}954 characters, removing 742{,}698{,}491 characters (47.3\%). Using the Qwen3-8B tokenizer, this corresponds to an estimated 220{,}166{,}861 fewer input tokens overall, or 220{,}166.9 fewer tokens per patient on average. 

\revision{Processing the sampled 1{,}000 timelines required 130.746 billable vCPU-hours, including file reading and writing. Assuming 8 GiB of memory per vCPU yielded an estimated 1{,}045.968 GiB-hours, corresponding to a total preprocessing cost of \$4.47. Extrapolation to approximately 3 million patients yielded 1.307 billion baseline notes and a one-time processing cost of \$13{,}403. Processing an additional 117.67 million notes annually yielded an estimated recurring cost of \$1{,}206 per year. After subtracting initial and annual preprocessing costs, projected net savings ranged from approximately \$70{,}368 to \$22.65 million over one year and from \$237{,}910 to \$67.96 million over three years depending on model and query patterns. Assuming one query per encounter, projected net savings ranged from approximately \$325{,}300 for Gemini 3.8 Flash to \$4.52 million for GPT-6 Astra over one year, and from \$1.00 million to \$13.58 million over three years depending on model choice (Supplementary Figure \ref{sfig:cost}).}

%% file: MAIN/030discuss.tex
We introduce TRACE, a method for reducing redundancy in clinical notes \revision{to support efficient downstream analysis}. TRACE leverages underutilized note attribution metadata embedded within electronic health record systems to identify templated and copied text. Across cohorts spanning hepatology, obstetrics, and inpatient care, as well as the MIMIC-III dataset, TRACE substantially reduced note length \revision{while retaining strong performance on information extraction and clinical outcome prediction tasks.} 
\revision{Importantly, economic analysis reveals substantial reductions in projected hospital-wide costs for LLM inference when using TRACE-processed notes compared to original notes. These results suggest the promise of note bloat removal algorithms like TRACE for improving the efficiency of LLMs for clinical applications. }
As health systems increasingly integrate generative AI into clinical workflows, approaches that improve the efficiency and structure of clinical text inputs may become an essential component of safely deploying these technologies at scale.

\revision{Unlike conventional text de-duplication approaches that rely on document-level similarity, TRACE uses attribution metadata to identify duplicated spans originating from templates or prior documentation, while aiming to preserve newly authored patient information.} This design enables detection of lightly edited copied text that would evade strict exact matching, as well as templated language that may occur infrequently within an individual patient record but is common across a broader patient population. By targeting documentation artifacts embedded within clinical data, TRACE provides a more \revision{targeted} approach to redundancy reduction than existing methods. More broadly, TRACE illustrates how underused provenance metadata embedded within electronic health record systems can be repurposed to improve the quality and efficiency of clinical text inputs for AI applications. This perspective aligns with a growing data-centric paradigm in which improving the quality and signal density of training data is as important as advances in model architecture \cite{zha2023datacentricai,zhang2024datacentricfmhealthcare}. 

Overall, we observe that \revision{the magnitude of} TRACE reductions varied across cohorts. Task-specific cohorts showed smaller character reductions (2\revision{4}–31\% when both modules could be applied) than the 47\% reduction observed in the mixed random 1{,}000-\revision{patient sample}, likely because the latter captured broader and more longitudinal documentation, \revision{ which provides more opportunities to identify repeated text. SHC’s role as a tertiary referral center may limit notes per patient in the cohort-specific datasets, reducing repeated text detectable by the Frequency Module. Metadata availability also influences TRACE’s ability to identify redundant text. In MIMIC-III, the Frequency Module alone reduced character volume by 12\%. While this modest reduction may reflect the absence of attribution metadata, which precluded use of the Reference Module, MIMIC-III's limited longitudinal and population coverage likely also constrains the Frequency Module’s detection of note bloat across patients, and a more complete reference corpus could enable greater reductions. Although attribution availability can impact TRACE text removal, the Reference Module requires only the note-level attribution tables available in Epic Clarity. Any institution with Epic Clarity access can therefore obtain the attribution data needed to run the Reference Module. This supports TRACE’s potential deployment across a wide range of settings, as Epic serves hospitals accounting for a majority and growing share of U.S. acute-care hospital beds (56.9\% in 2025) \cite{bruce_epic_market_share_2026}. These findings highlight the value of preserving provenance information in healthcare datasets to support more efficient downstream analysis.} 

Reducing redundant clinical documentation has important implications for the scalability of AI systems in healthcare. \revision{While TRACE cost savings are relatively small on the per-patient, per-query level (\$0.17 to \$2.20 per full-timeline query), this translates to considerable savings at the institutional level. Under the assumption of just one query per encounter, estimated net savings over three years range from approximately \$1.00 million with Gemini 3.8 Flash to \$13.58 million with GPT-6 Astra, after accounting for initial and annual TRACE preprocessing costs.} Similarly, latency gains may be small on the individual level, but even 100 milliseconds saved per \revision{query} could add up to as much as 57 hours of clinician time saved over the course of a year. Context space is also at a premium in clinical AI workflows, particularly when clinical notes must be considered alongside external sources of information such as practice guidelines, retrieved biomedical literature, institutional protocols, or care pathways. By reducing note redundancy, TRACE can free capacity within fixed context windows for these complementary inputs and may make higher-capacity models practical in settings where they would otherwise be financially prohibitive. Beyond inference, removal of duplicated text may also improve training efficiency when models are trained or fine-tuned on institutional data. Large-scale de-duplication of common NLP corpora has been shown to improve training efficiency, requiring fewer training steps to achieve the same (or better) accuracy \cite{lee-etal-2022-deduplicating}. 

Beyond financial and computational benefits, reducing site-specific documentation artifacts may improve the robustness of clinical AI models. With the widespread prevalence of templates and automatically-imported structured data, models learn spurious associations during training that fail to generalize across institutions and temporal contexts. This phenomenon, termed ``shortcut learning'' \cite{geirhos_shortcut_2020}, has also been documented in clinical applications, where imaging models have falsely associated demographic or acquisition-related features with disease labels \cite{yang_limits_2024,ongly_shortcut_2024,hill_shortcutting_2024} or LLMs have over-indexed on specific topics in notes \cite{sakib_spurious_2025}. Templated information is also subject to shifts in documentation practices across sites, specialties, or time; removing these elements at scale may improve generalizability and robustness to these shifts.

This study has several limitations \revision{that motivate future work. First, evaluation of templated and copied text was constrained by the available reference data.} Although we rigorously evaluate templated text identification, the Clarity database does not provide character-level ground truth specifying the exact spans imported during copy–paste events, limiting direct evaluation of copied text detection. \revision{Moreover, because the gold-standard annotations were derived from the same Clarity-linked templates used by the Reference Module, they capture only attribution-linked templated content, and recall estimates may be optimistic. In addition, TRACE currently does not explicitly identify individual smart elements. The Frequency Module may remove repeated but clinically relevant findings in semi-structured elements such as Physical Exam macros, while partially edited elements may retain redundant text that escapes detection. Our broader objective was to assess whether TRACE could substantially reduce text volume while maintaining downstream task performance, rather than exhaustively identify all templated content. Metadata availability and documentation structure also affect TRACE's applicability. Further evaluation across institutions, note types, inference and embedding models, context-window sizes, and clinical settings is needed to characterize TRACE performance in a wider range of settings.} \revision{Finally, statistical comparisons of downstream performance also require qualification. TRACE substantially reduced input text, with small changes in aggregated information extraction and clinical outcome prediction performance that met post-hoc non-inferiority criteria, supporting its potential to improve clinical NLP efficiency. However, task-specific sensitivity–specificity trade-offs warrant consideration for individual applications. }  

Several directions for future research could extend the impact of TRACE. A key next step is to determine whether training on TRACE-processed notes improves sample efficiency, reduces reliance on duplicated documentation artifacts, and strengthens out-of-distribution performance, including cross-site generalization. TRACE could also be expanded to target additional sources of redundancy embedded within narrative text, such as auto-populated structured elements including laboratory values, vital signs, and other numeric data, potentially enabling further compression and more efficient integration of structured and unstructured information. More broadly, studying TRACE across additional institutions, patient populations, and downstream tasks will help clarify when redundancy reduction most improves the efficiency, robustness, and scalability of clinical AI systems.\revision{ From a clinical standpoint, future work will also explore whether TRACE-processed notes can facilitate more efficient chart review for human clinicians. Possible applications include alternative note views with templated and copied content reduced, as well as modified search functions that use the fraction of text retained after TRACE processing to help prioritize search results.}

\subsection*{Conclusion}
TRACE is a lightweight, open-source method for pre-processing clinical notes that can be integrated into existing AI workflows to reduce documentation redundancy, compress longitudinal patient histories, and lower token-driven inference costs. Across diverse cohorts and tasks, TRACE achieves substantial compression\revision{ while maintaining overall performance}. By showing that underused EHR metadata can be leveraged to improve the quality and efficiency of clinical text inputs, this work highlights note pre-processing as an important infrastructure layer for scalable deployment of clinical AI.

%% file: MAIN/050figures.tex

\begin{figure}[h!]
\centering
\includegraphics[width=\textwidth
]{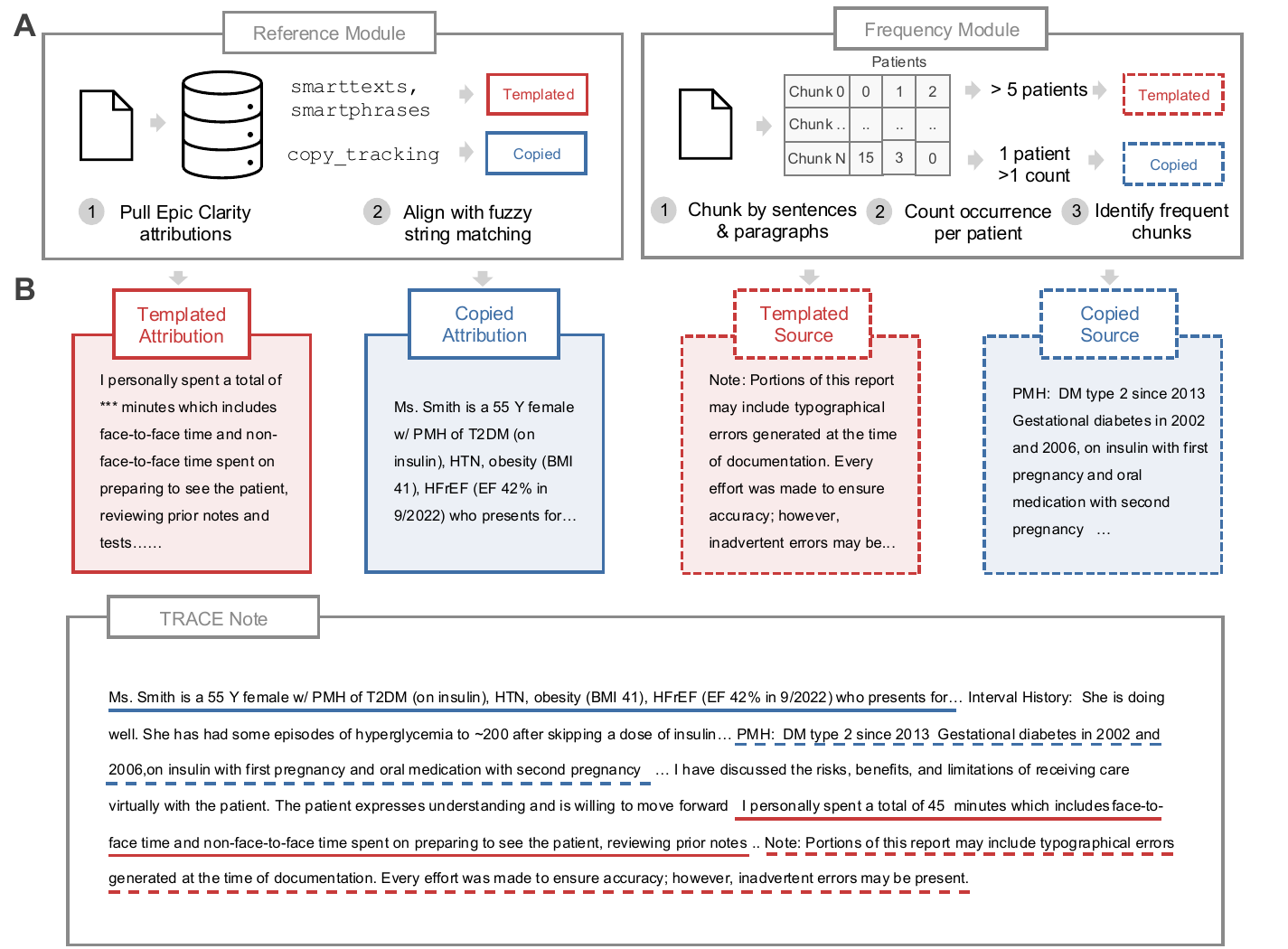}
\caption[]{\textbf{TRACE Method.} (A) TRACE identifies templated and copied text using Epic Clarity attribution metadata (Reference Module) and flags additional redundant text blocks using frequency-based de-duplication (Frequency Module).
(B)\revision{ Example TRACE-annotated note showing templated and copied spans with sources identified by the Reference and Frequency Modules.}
}
\label{fig:method} 
\end{figure}

\begin{figure}[h!]
\centering

\includegraphics[width=\textwidth]{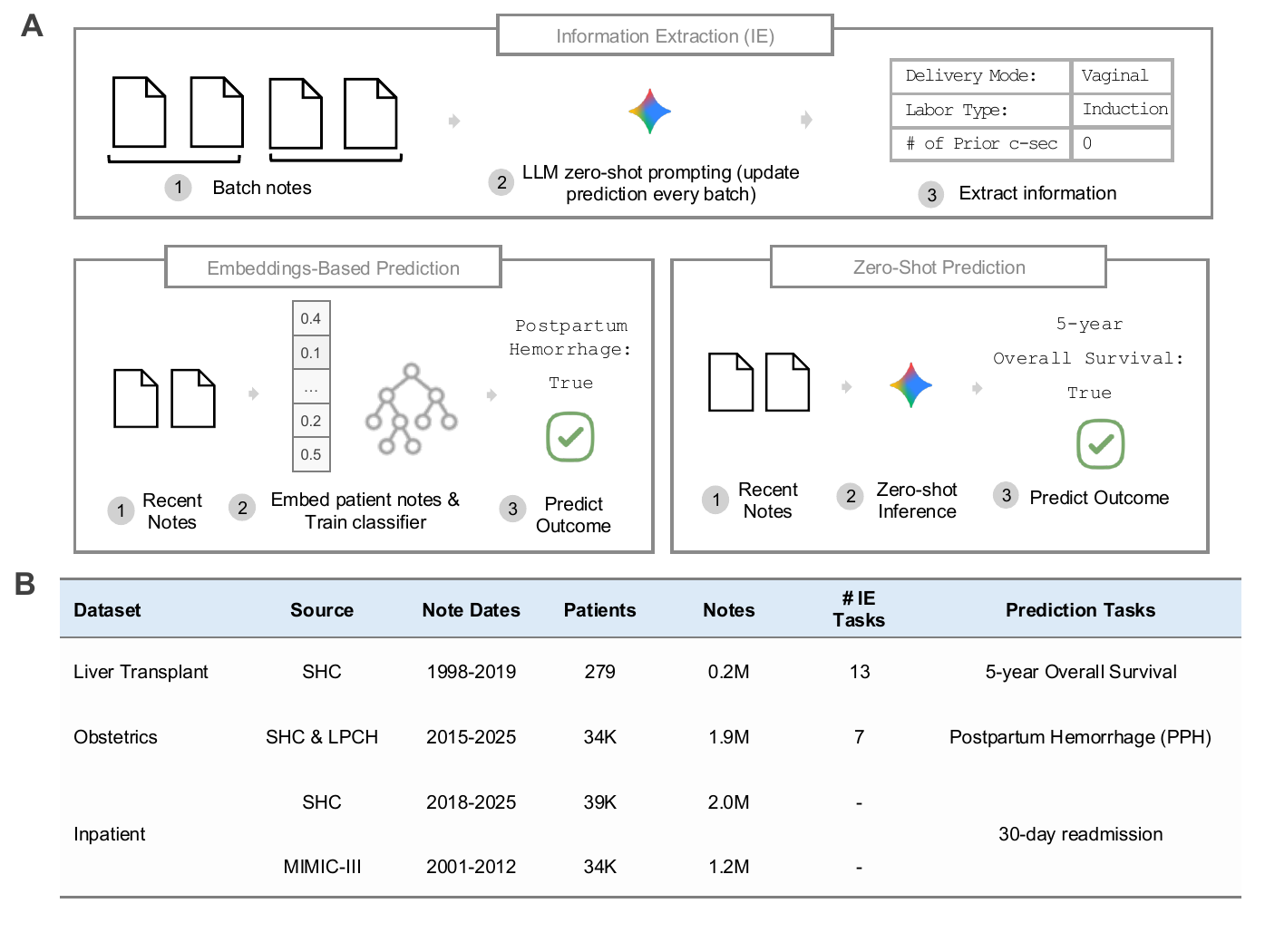}

\caption[]{\textbf{TRACE downstream tasks.} (A) TRACE utility is evaluated through two downstream tasks: Information Extraction (IE) and Clinical Outcome Prediction. In IE, a large language model performs inference over batched notes and updates the prediction as new notes are added. In clinical outcome prediction, \revision{the most recent 32K tokens (embedding-based) or up to 1M tokens (zero-shot) are used}. Notes can either be used for zero-shot inference or embedded and used to train a classifier. (B) Datasets for downstream evaluation.}
\label{fig:validation} 
\end{figure}

\begin{figure}[h!]
\centering

\includegraphics[width=\textwidth]{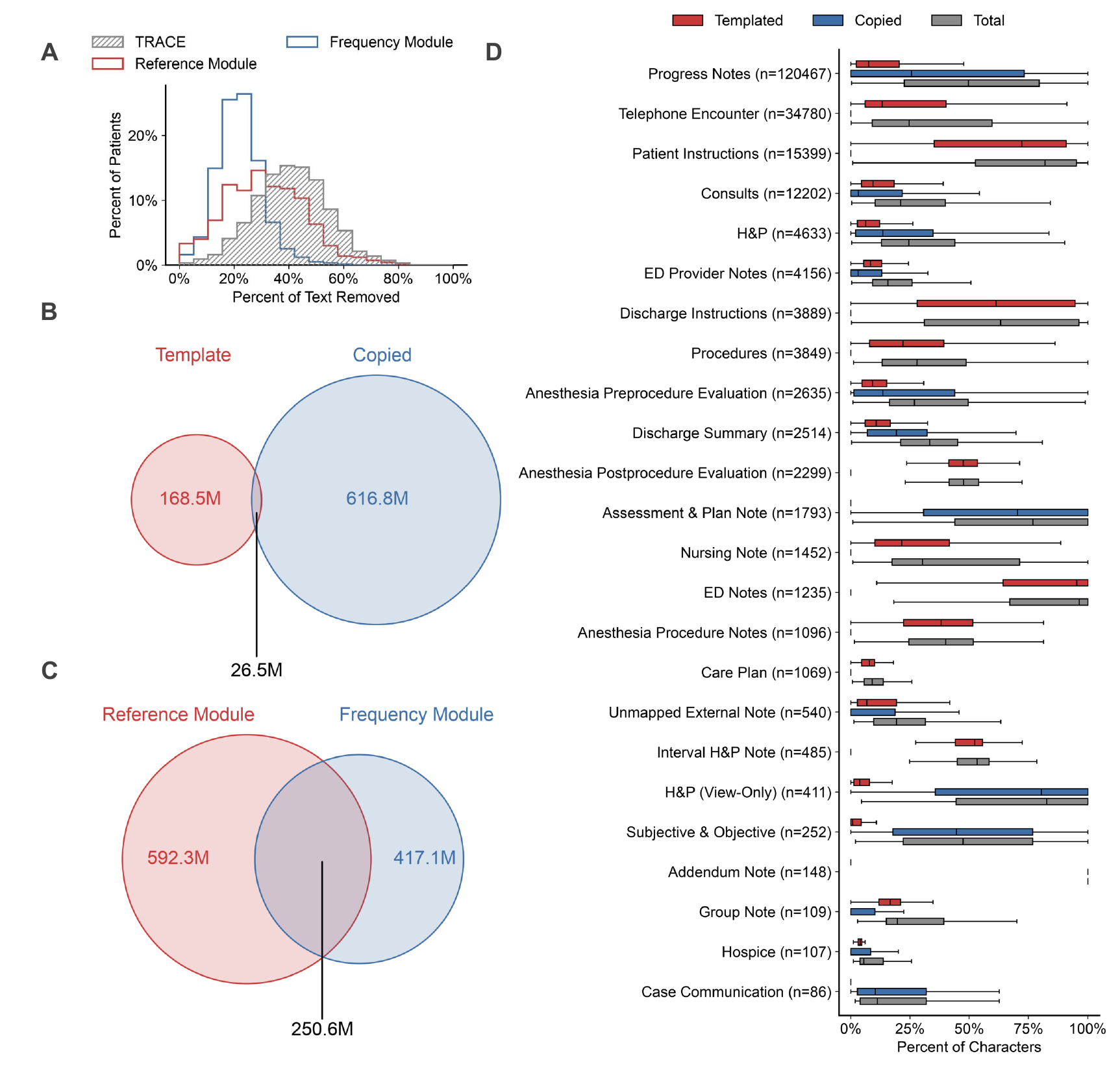}
\caption[]{\textbf{TRACE facilitates the analysis of templated and copied text in 1{,}000 randomly sampled patients} (A)  Distribution of text removal by TRACE on full note timelines. (B) Number of characters identified as Templated, Copied, and the intersection. (C) Number of characters identified with Reference and Frequency Modules. \revision{Counts reflect flagged characters, while the intersection refers to characters identified by both methods. Because copied spans are retained at their first occurrence, flagged characters exceed removed characters.} (D) Percent of templated and copied text for the 24 most common note types in Stanford Health Care.}
\label{fig:random} 
\end{figure}

\begin{figure}[h!]
\centering

\includegraphics[width=\textwidth]{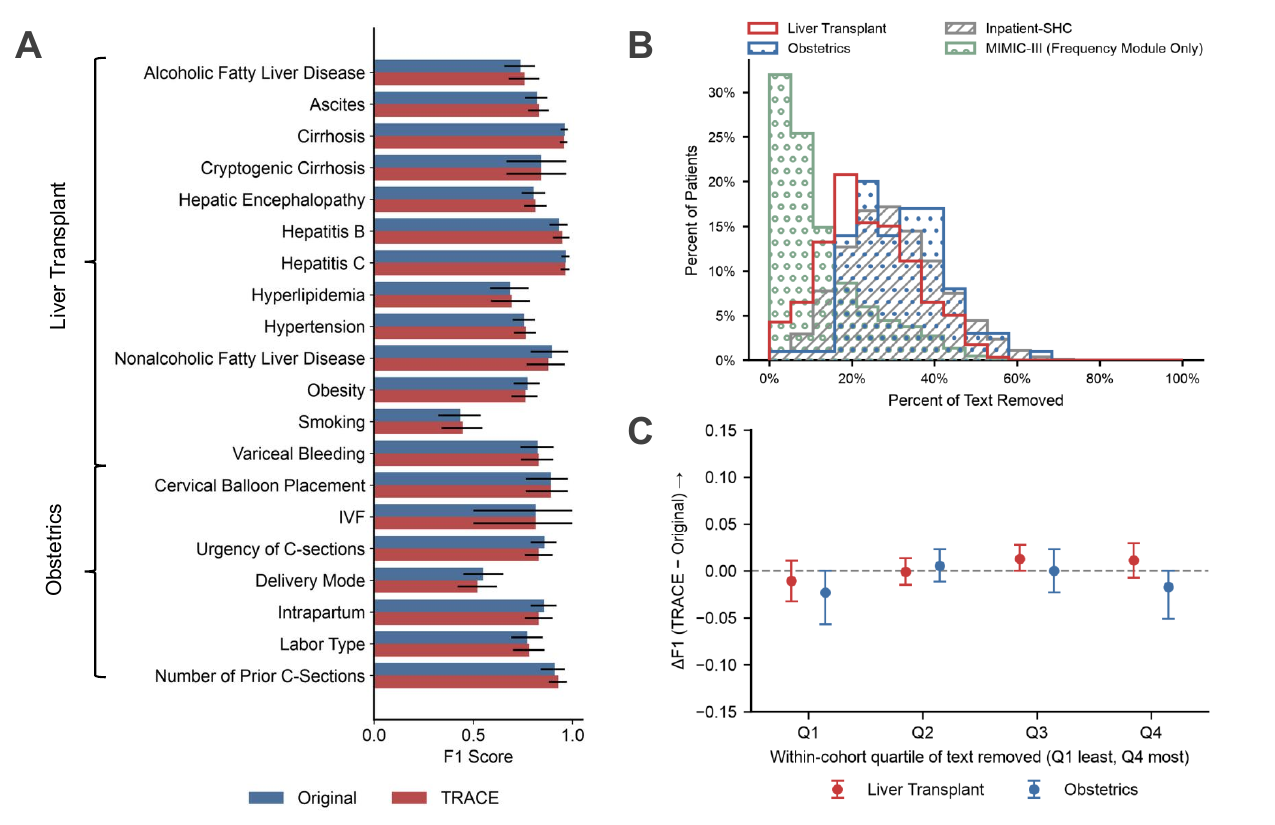}

\caption[]{\textbf{\revision{TRACE-processed notes achieve non-inferior performance in information extraction (IE) compared with original notes}} (A) IE performance across tasks in hepatology and obstetrics for original and TRACE Notes. The X axis indicates the F1 score for the respective cohort. The Y axis indicates the specific task. (B) Distribution of text removed in patient timelines across four real world cohorts. (C) \revision{Paired F1 differences (TRACE -- Original), averaged across IE tasks, by within-cohort quartile of the percentage of characters removed by TRACE (Q1, least; Q4, most). Points show observed differences; error bars show 95\% percentile bootstrap confidence intervals (1{,}000 replicates).}}
\label{fig:ie} 
\end{figure}

\begin{figure}[h!]
\centering

\includegraphics[width=\textwidth]{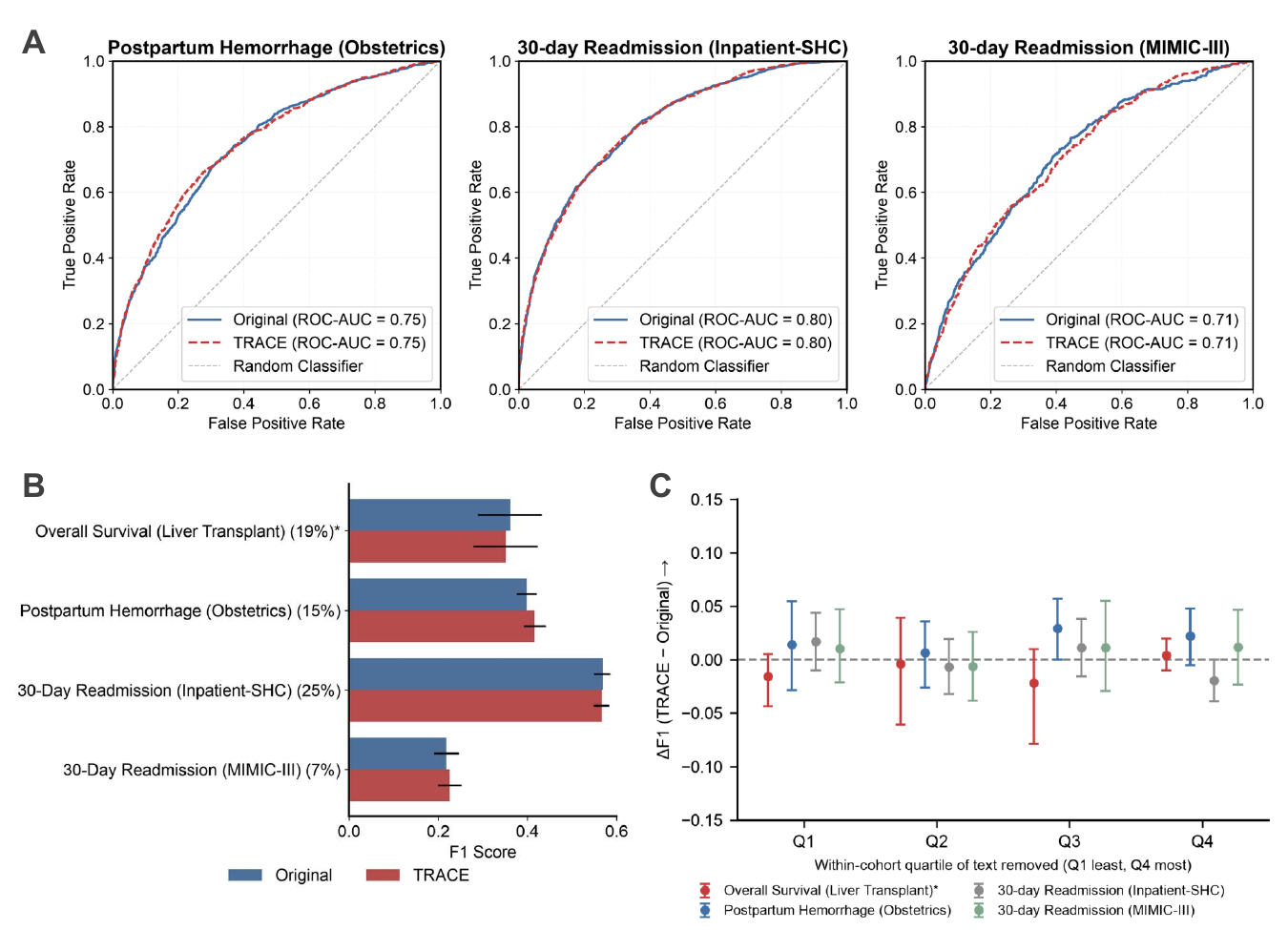}
\caption[]{\textbf{TRACE enables efficient clinical outcome prediction.} A) \revision{ROC} curves for Postpartum Hemorrhage and 30-day Readmission B) F1 scores with 95\% bootstrapped confidence intervals for zero-shot Overall Survival prediction (*) and embedding-based PPH and readmission prediction C)\revision{ Paired F1 differences (TRACE minus Original) by within-cohort quartile of the percentage of characters removed by TRACE (Q1, least; Q4, most). Points show observed differences; error bars show 95\% percentile bootstrap confidence intervals (1{,}000 replicates). }}
\label{fig:outcome} 
\end{figure}

\restoregeometry

%% file: SI/SI-notes.tex
\vspace{1em}

\subsection*{Supplementary Note 1: TRACE Runtime Analysis}
\label{note:runtime}
TRACE removes redundant clinical note text in a targeted manner to improve efficiency. The Reference Module leverages note attributions to enable high-precision span removal with best-case linear time and worst-case quadratic time in the number of attributed spans per note. The Frequency Module scales linearly with the number of note chunks and can be applied when attribution data are unavailable. In practice, these complexity guarantees make TRACE scalable across large cohorts. The Reference Module required approximately 12 hours per 50,000 notes on a single CPU with 8GiB RAM ($\approx$ 4,167 notes/hour). This CPU-centered framework can easily scale to large datasets. The Frequency Module compute varied across cohorts. The Inpatient-SHC cohort (2M notes) was processed in 1 hour on a single CPU with 64GiB RAM, while smaller cohorts (0.2M notes) required $\approx$ 15 minutes on 32GiB RAM.

\subsection*{Supplementary Note 2: Inpatient-SHC Readmission Labels}
\label{note:readmission}
Index admissions were defined as encounters that included at least one discharge summary. Death-related encounters were excluded by identifying the longest discharge summary containing at least one term from each of the following two groups: (``discharge summary'', ``admission'', ``discharge'', or ``d/c'') and (``death certificate'', ``death pronouncement'', ``death worksheet'' or ``time of death.'')

\subsection*{\revision{Supplementary Note 3: TRACE Minimum Span Ablation}}
\label{note:ablation}

\revision{To assess sensitivity to the minimum span length, we repeated TRACE with thresholds of 25, 50, and 75 characters, applied separately within each module as described in the Methods. In the Reference Module, spans separated only by whitespace were merged, and spans shorter than the threshold were discarded. In the Frequency Module, normalized chunks with fewer non-space characters than the threshold were excluded before frequency counting. For each threshold, we measured corpus-level and mean per-timeline text removal in the 1{,}000-patient sample (Supplementary Table~\ref{tab:ablation_removal}), template-detection precision and recall in the 100-note annotated subset (Supplementary Table~\ref{tab:ablation_accuracy}), and embedding-based PPH prediction (Supplementary Table~\ref{tab:ablation_pph}). Recall was evaluated using the same reference annotations and eligible reference spans ($\geq$50 characters) at every threshold to ensure comparison against a fixed reference set. Downstream ablation was limited to PPH because of the computational cost of repeating downstream experiments.}

%% file: SI/SI-figures.tex

\begin{figure*}[p]
\centering
\includegraphics[
    width=\textwidth,
    height=0.97\textheight,
    keepaspectratio,
    trim=0 15pt 0 12.5pt,
    clip
]{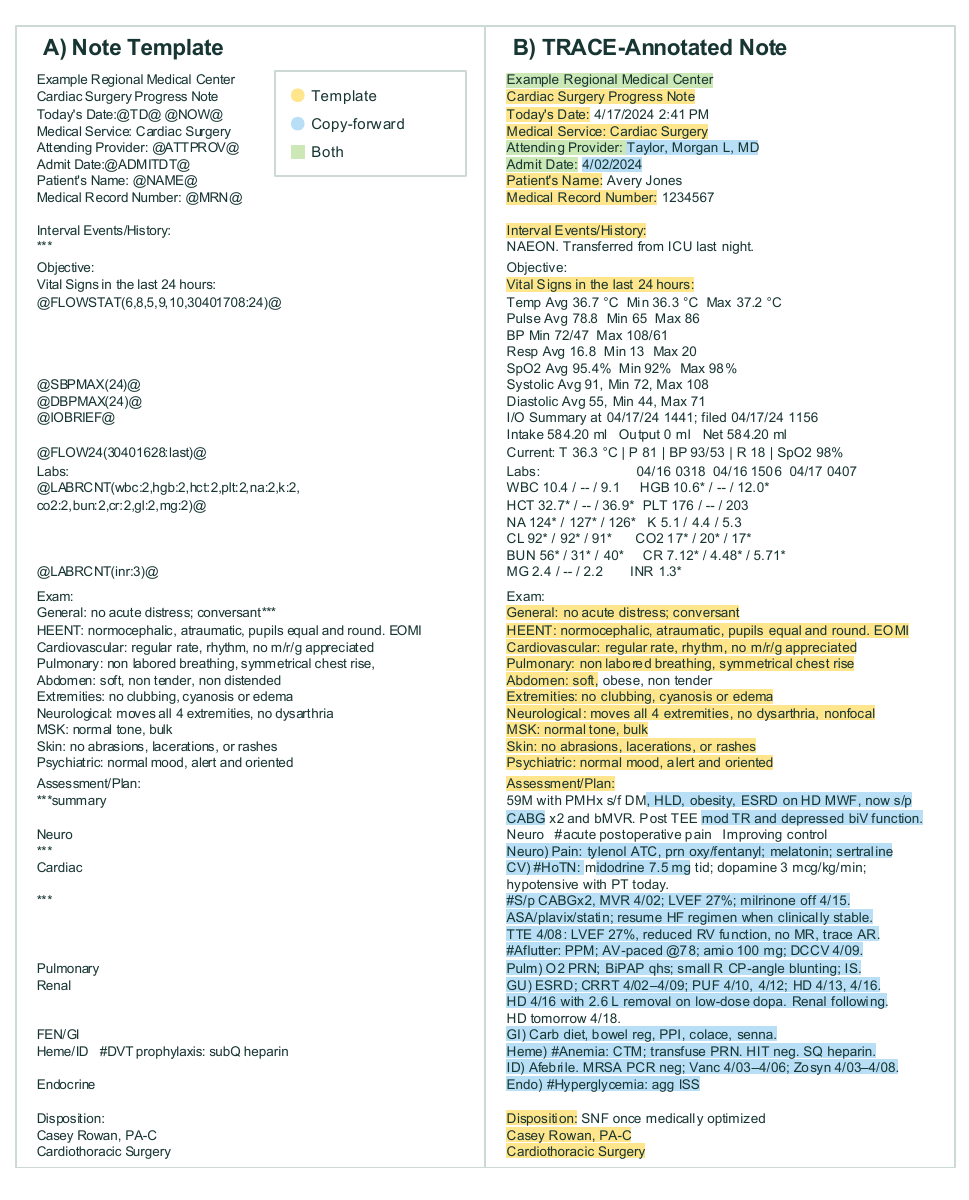}
\captionsetup{skip=2pt}
\caption{\revision{\textbf{Example of TRACE span identification.}
(A) Source template. (B) Corresponding de-identified
note with templated, copy-forward, and overlapping spans highlighted. Edits have been made to ensure de-identification, changing some wording and minor information while retaining identified spans. 
Copy-forward spans are retained at their first occurrence.
Highlights show identified spans, while removal depends on module-specific length thresholds.}}
\label{sfig:note_example}
\end{figure*}

\begin{figure*}
\centering
\small
\begin{tcolorbox}[colback=gray!5, colframe=gray!40!black, title=Span Drafting Prompt Part 1, width=\textwidth, boxrule=0.5pt, arc=2mm
]
You are a clinical annotator. Your goal is to highlight the parts of a clinical note that came from 1) templated text, or 2) structured data (labs, vitals, demographics, meds, or other info in the patient's chart).

\begin{enumerate}
    \item Identify sections of the full clinical note that are likely from the template(s) (identifying matching text) and identify them with the following start/stop characters [[T]] text here [[/T]].
    \item Identify sections of the note that are likely from structured data (often denoted with the @ symbol as part of smartlinks: for example, @MEDLIST@ will populate a full list of patient meds). Highlight these with [[S]] for start and [[/S]] for stop at the end of the text that comes from the smartlink.
    \item Only return the VERBATIM text of the note; where it differs with the template text only return the note text itself. The only different characters between the original and the returned version should be the template annotations.
\end{enumerate}
Remember:
\begin{enumerate}
    \item Close every span that's opened and avoid overlapping spans
    \item The user may have changed or edited the template. Do your best to determine which text came from the template itself by identifying identical spans of text, even if they are interspersed with new text.
    \item If a structured span is a smartlink (@) in the template, the corresponding text should be marked as structured [[S]] in the highlights. Sometimes it is difficult to tell what text came from a smartlink; you can do your best based on the location of the @TEXT@ in the template and the surrounding text in the template/note.
    \item Do not change any of the final note characters - return them verbatim with the highlighting characters
    \item Match text from ANY of the provided templates
\end{enumerate}

``Labs drawn and reviewed." does NOT match ``Labs drawn 1/15 reviewed."
``Patient denies" does NOT match ``Pt denies." When unsure, DON'T highlight it. The templated text you are given may not actually be present in the note. If this is the case, just return an empty string.
\end{tcolorbox}

\caption{\revision{Prompt used with GPT-4.1 to draft candidate templated and structured-data spans for the metadata-assisted reference annotations.}}
\label{sfig:prompt_spans}
\end{figure*}

\begin{figure*}
\centering
\begin{tcolorbox}[colback=gray!5, colframe=gray!40!black, title=Information Extraction System Prompt, width=\textwidth, boxrule=0.5pt, arc=2mm
]
You are a clinical information extraction assistant.

Your task is to carefully and comprehensively review patient clinical notes and determine whether there is sufficient evidence that the patient has each specified medical condition.

Rules:
\begin{enumerate}
    \item Base your determinations ONLY on information explicitly stated or strongly implied in the notes.
    \item Do NOT infer diagnoses from weak, ambiguous, or unrelated evidence.
    \item Do not add conditions not provided.
    \item Do not include explanations, comments, or text outside the requested JSON.
\end{enumerate}
These are the variables you will extract. Return a choice for each variable:
\{VARIABLES\}

Output must be valid JSON. This is the only valid schema:
\{SCHEMA\}

\end{tcolorbox}

\caption{Prompt used for information extraction for the first batch}
\label{sfig:prompt_ie}
\end{figure*}

\begin{figure*}
\centering
\begin{tcolorbox}[colback=gray!5, colframe=gray!40!black, title=Information Extraction Update System Prompt, width=\textwidth, boxrule=0.5pt, arc=2mm
]
You are a clinical information extraction assistant operating in update mode.

Your task is to review newly provided patient notes and update an existing condition-assessment JSON accordingly.

Rules:
\begin{enumerate}
    \item Start from the existing JSON assessment as the baseline.
    \item Review the new notes, recorded at a later date, carefully and comprehensively.
    \item  More recent notes take precedence over older information.
    \item Change a condition’s value ONLY if the new notes provide clear, sufficient evidence to do so.
    \item If the new notes do not affect a condition, leave its value unchanged.
    \item Do not add or remove conditions.
    \item Do not include explanations, comments, or text outside the requested JSON.
\end{enumerate}
These are the variables you will extract. Return a choice for each variable:
\{VARIABLES\}

Output must be valid JSON. This is the only valid schema:
\{SCHEMA\}

\end{tcolorbox}

\caption{Prompt used to update extracted values after the first batch}
\label{sfig:prompt_ie_update}
\end{figure*}

\begin{figure*}
\centering
\begin{tcolorbox}[colback=gray!5, colframe=gray!40!black, title=Outcome Prediction System Prompt, width=\textwidth, boxrule=0.5pt, arc=2mm
]
You are a clinical prediction assistant.

Your task is to carefully review patient clinical notes and predict if the patient will experience the medical outcome in the future based on the current patient records.

Rules:
\begin{enumerate}
    \item Do NOT infer outcome from weak, ambiguous, or unrelated evidence.
    \item Do not add outcomes not provided.
    \item Do not include explanations, comments, or text outside the requested JSON.
\end{enumerate} 
These are the outcomes you will predict. Return a choice for each outcome:
\{OUTCOMES\}

Output must be valid JSON. This is the only valid schema:
\{SCHEMA\}

\end{tcolorbox}

\caption{Prompt used for zero-shot prediction of outcomes.}
\label{sfig:prompt_zero}
\end{figure*}

\input{SI/FIG/gold_span_bootstrap}

\input{SI/FIG/ablation}
\input{SI/FIG/notetype_total}

\input{SI/FIG/notetype_template}
\input{SI/FIG/notetype_copyforward}

\input{SI/FIG/performance}
\begin{figure}[h!]
\centering
\includegraphics[width=\textwidth]{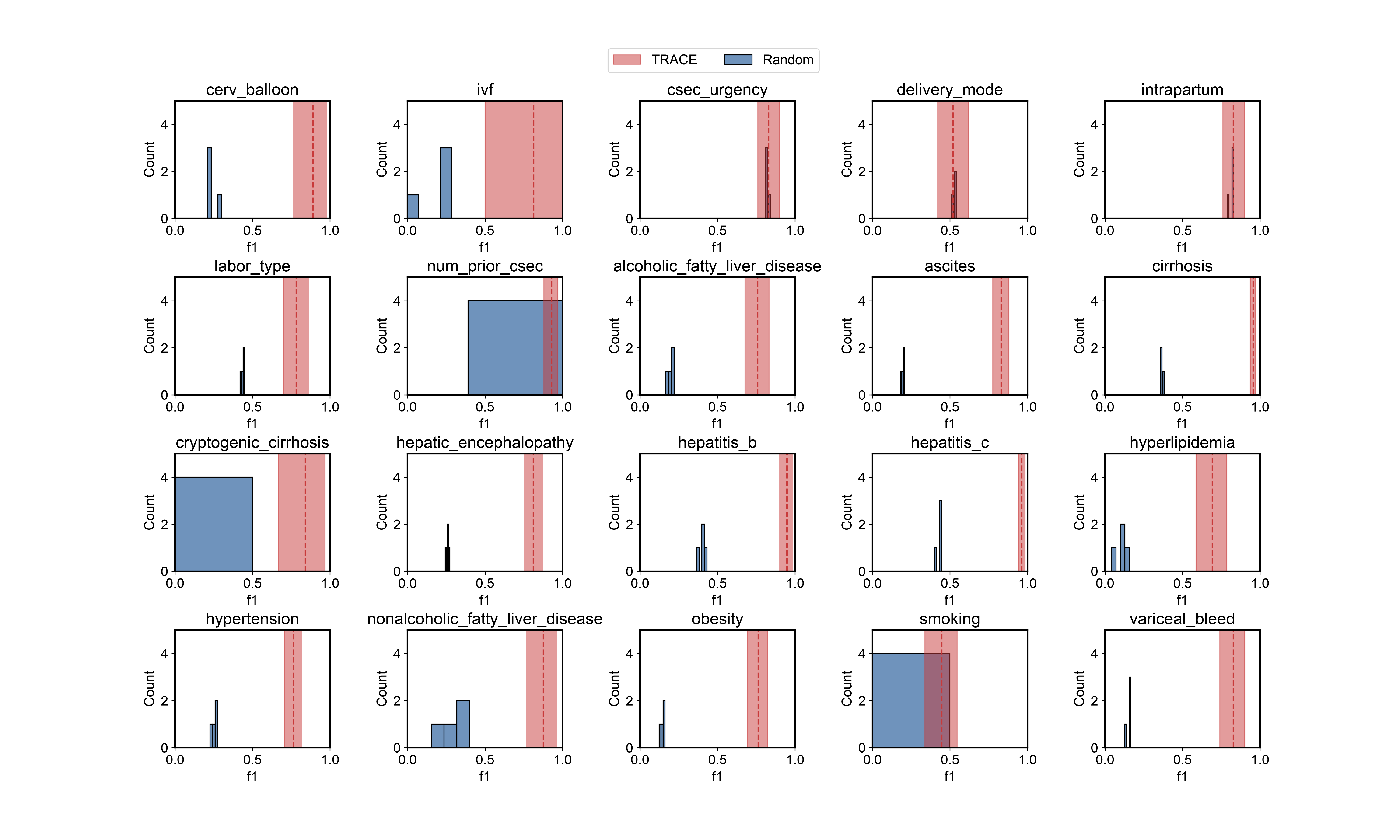}
\caption[]{TRACE 95\% confidence interval for Information Extraction F1 compared to distribution of F1 for random removal}
\label{sfig:random}
\end{figure}

\input{SI/FIG/logistic_arm_coefficients}
\input{SI/FIG/logistic_spline_tests}

\begin{figure}
\includegraphics[width=\textwidth]{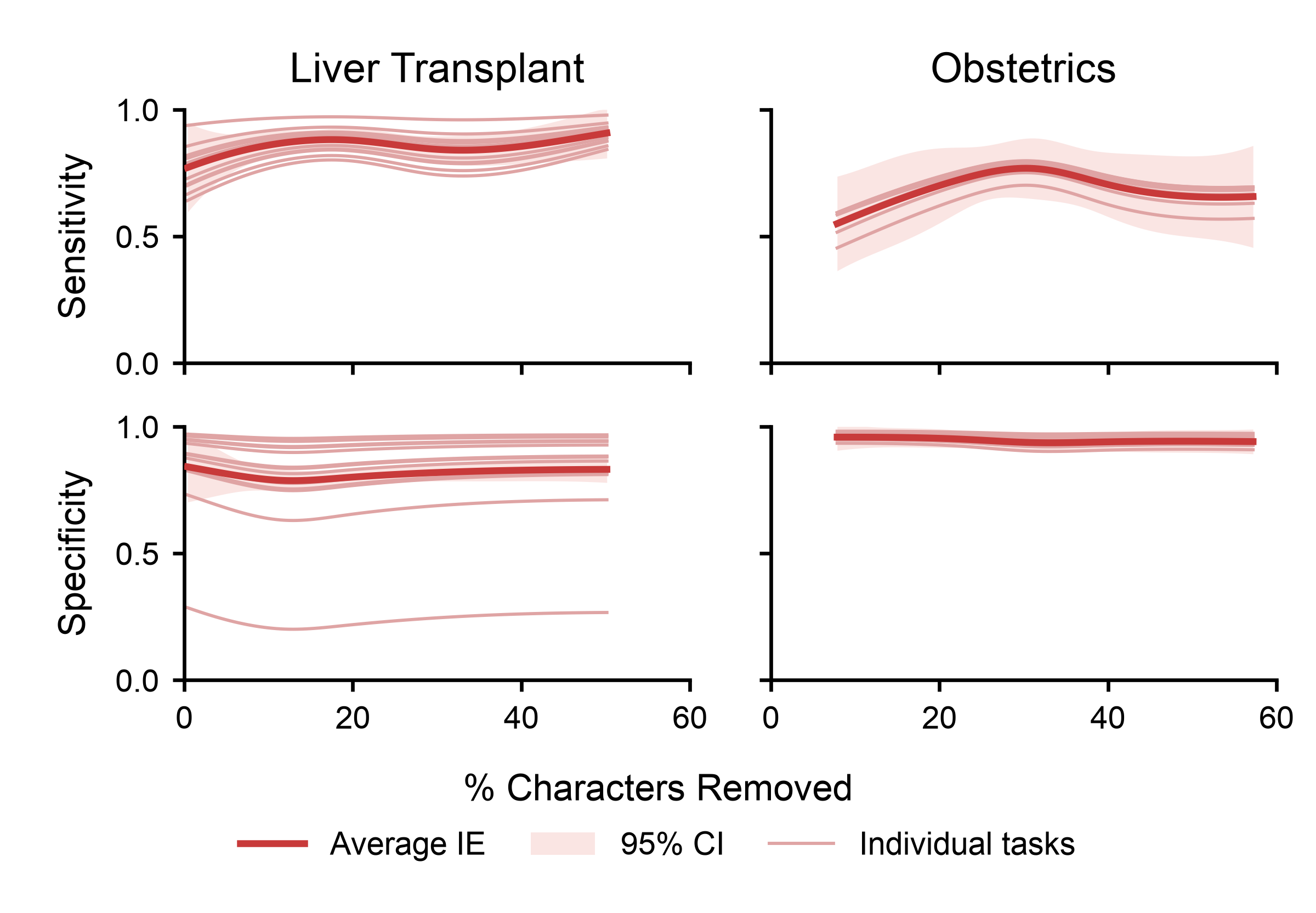}
\caption[]{\revision{\textbf{Association between text removal and information extraction accuracy.} Adjusted predicted probability of a correct prediction among TRACE-processed inputs as a function of the percentage of characters removed by TRACE, estimated from restricted cubic spline logistic regression models (Equation 3) fit separately among reference-positive (sensitivity) and reference-negative (specificity) observations. Original input length was held at its mean and feature prevalence at the task-specific prevalence. Thick lines show the average across IE tasks; thin lines show individual tasks. Shading denotes pointwise 95\% confidence intervals from the patient-cluster-robust covariance matrix.}}
\label{sfig:spline-ie}
\end{figure}

\begin{figure}
\includegraphics[width=\textwidth]{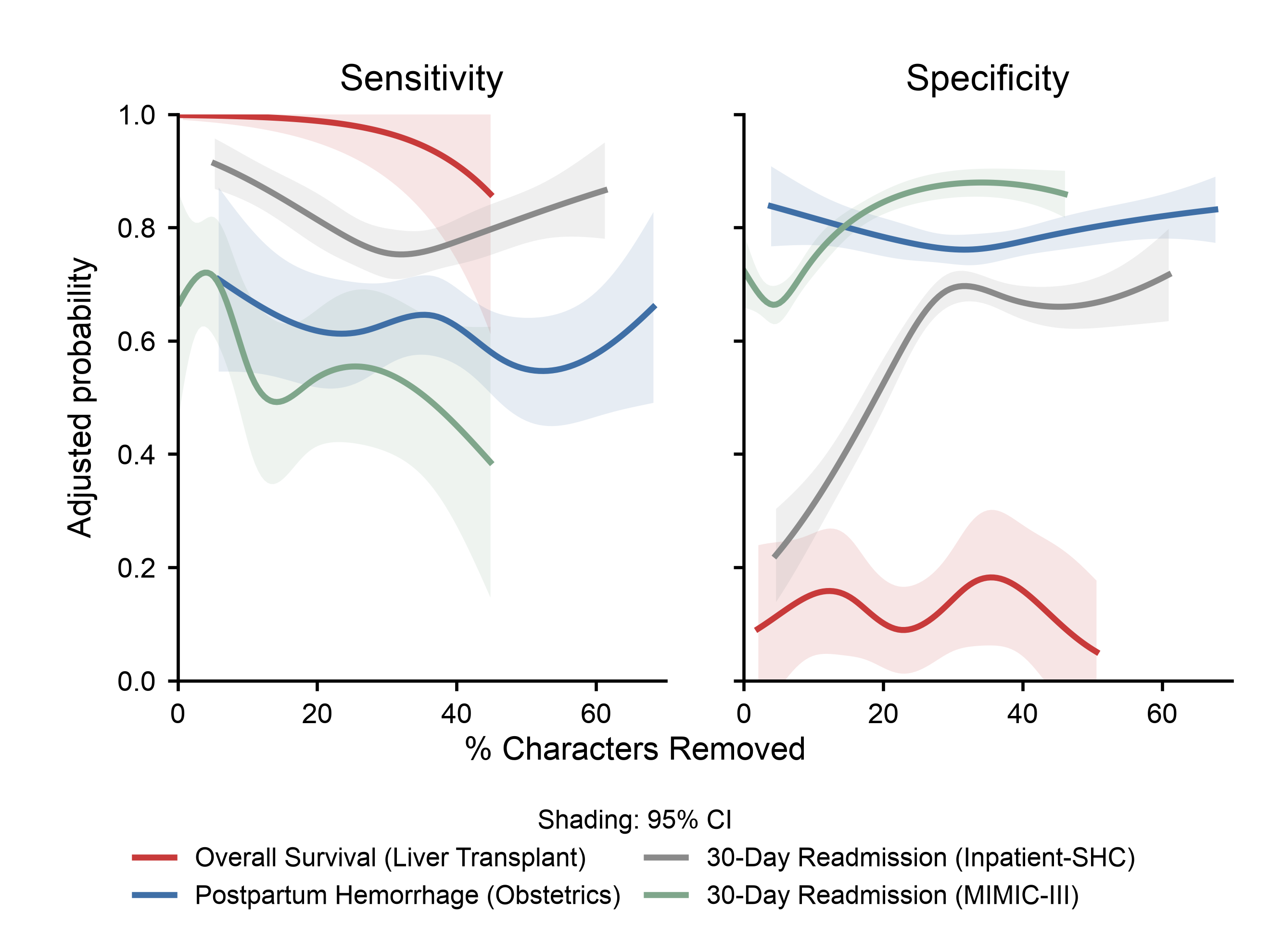}
\caption[]{\revision{\textbf{Association between text removal and clinical outcome prediction accuracy.} Adjusted predicted probability of a correct prediction among TRACE-processed inputs as a function of the percentage of characters removed by TRACE, estimated from restricted cubic spline logistic regression models (Equation 3) fit separately among reference-positive (sensitivity) and reference-negative (specificity) observations, with original input length held at its mean. Shading denotes pointwise 95\% confidence intervals from the patient-cluster-robust covariance matrix. Overall survival sensitivity uses a linear term for text removal because the spline model was singular.}
\label{sfig:spline-pred}}
\end{figure}

\input{SI/FIG/noninferiority_mht}

\input{SI/FIG/out_bootstrap_p}
\input{SI/FIG/out_bootstrap_r}
\input{SI/FIG/out_bootstrap_f}
\input{SI/FIG/out_bootstrap_auroc}
\input{SI/FIG/outcome_auprc}
\input{SI/FIG/ablation_pph}


\begin{figure}[h!]
\centering
\includegraphics[width=\textwidth]{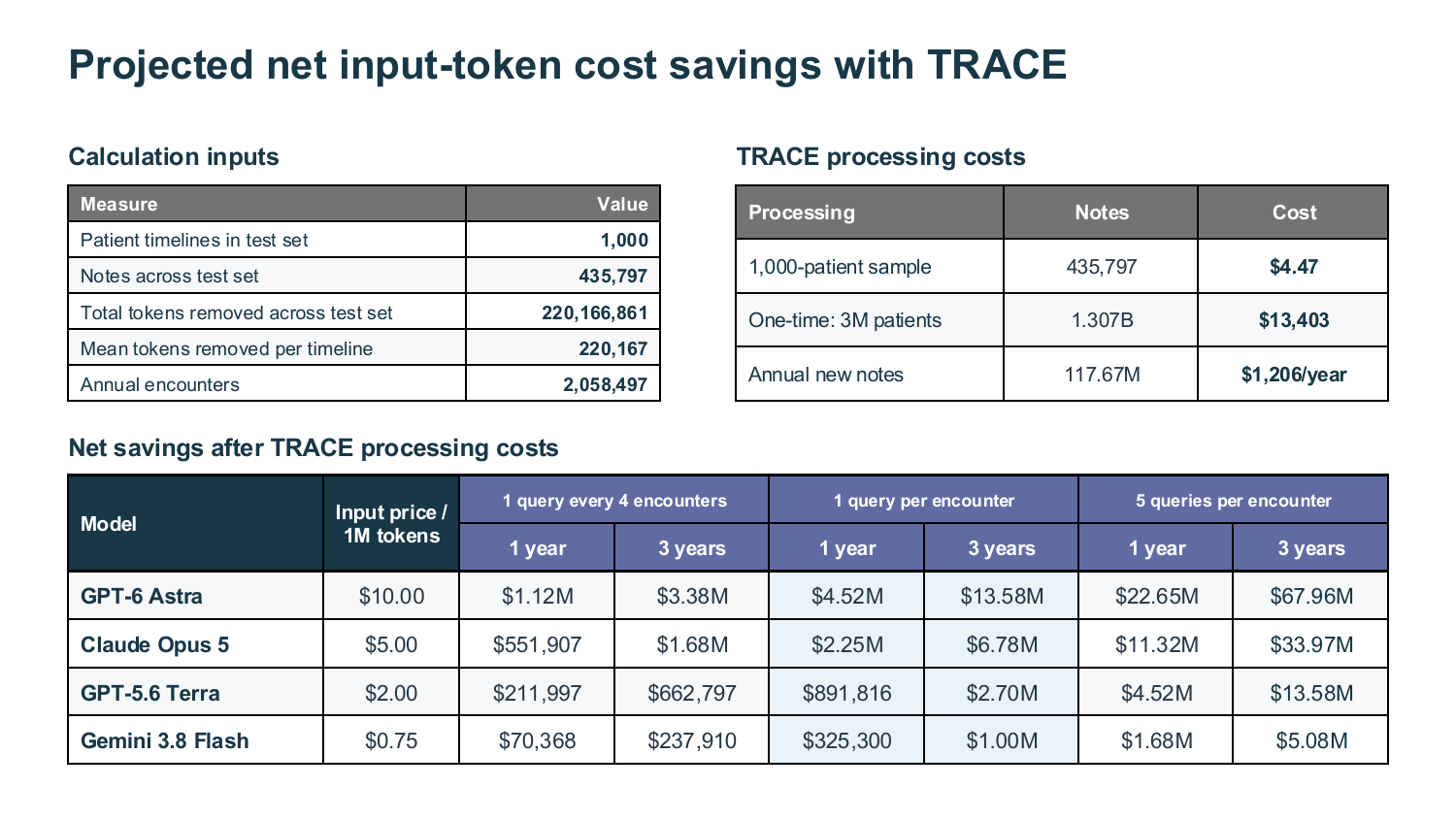}
\caption[]{\revision{Estimated net system-wide cost savings with TRACE. Gross annual savings were estimated from the mean reduction of 220,167 tokens per patient timeline (based on the 1{,}000-patient timeline sample), 2{,}058{,}497 FY2024 Stanford Health Care encounters (from published utilization data), and the model prices and query frequencies shown. Each query was assumed to process a full timeline. Processing 435,797 notes from 1,000 patients required 130.746 billable vCPU-hours. Assuming 8 GiB per vCPU gives 1,045.968 GiB-hours and a sample processing cost of \$4.47 at Google Cloud E2 virtual machine 1-year committed-use rates under an existing institutional commitment. Extrapolation to 3 million SHC patients gives 1.307 billion baseline notes and an initial cost of \$13,403. New notes equal to a fixed 9\% of baseline volume (approximately 117 million notes) adds a fixed cost of \$1,206 annually, assuming stable model pricing. Net savings subtract initial and annual processing costs beginning in year 1. Projections assume representative sample timelines, constant average processing cost per note, incremental processing without rerunning the historical corpus, and constant query volume and token reduction. Storage and other operational costs are excluded.

}}
\label{sfig:cost}
\end{figure}

%% file: SI/FIG/gold_span_bootstrap.tex
\begin{table}[htbp]
    \centering
    \small
    \setlength{\tabcolsep}{5pt}
    \sisetup{table-format=2.2, detect-weight, mode=match}

    \begin{tabular}{@{}ll SSSS@{}}
        \toprule
        & & & & {\textbf{Written by}} & \\
        \textbf{Method} & & {\textbf{Structured}} & {\textbf{Template}}
            & {\textbf{note author}} & {\textbf{Indeterminate}} \\
        \midrule

        \multicolumn{6}{@{}l}{\textbf{Annotator 1}} \\
        \addlinespace[2pt]
        \textbf{TRACE} & \textit{Reference}
            & 0.06 & \bfseries 91.27 & 1.96 & 6.71 \\
        & & {\footnotesize(0.00--0.18)} & {\footnotesize(85.58--95.69)}
            & {\footnotesize(0.57--4.06)} & {\footnotesize(2.87--11.82)} \\
        \addlinespace[3pt]
        & \textit{Frequency}
            & 3.08 & \bfseries 90.11 & 0.29 & 6.52 \\
        & & {\footnotesize(1.05--6.13)} & {\footnotesize(82.93--94.70)}
            & {\footnotesize(0.00--0.93)} & {\footnotesize(2.86--12.07)} \\
        \addlinespace[5pt]
        \textbf{Random} &
            & \bfseries 41.59 & 30.68 & 14.87 & 12.86 \\
        & & {\footnotesize(30.66--51.30)} & {\footnotesize(21.94--41.10)}
            & {\footnotesize(10.39--19.90)} & {\footnotesize(8.01--18.35)} \\

        \midrule

        \multicolumn{6}{@{}l}{\textbf{Annotator 2}} \\
        \addlinespace[2pt]
        \textbf{TRACE} & \textit{Reference}
            & 2.32 & \bfseries 77.11 & 6.61 & 13.96 \\
        & & {\footnotesize(0.32--6.04)} & {\footnotesize(65.72--85.99)}
            & {\footnotesize(2.30--12.54)} & {\footnotesize(7.71--23.11)} \\
        \addlinespace[3pt]
        & \textit{Frequency}
            & 1.80 & \bfseries 79.67 & 2.63 & 15.90 \\
        & & {\footnotesize(0.41--4.14)} & {\footnotesize(66.44--88.24)}
            & {\footnotesize(0.86--5.55)} & {\footnotesize(8.28--27.36)} \\
        \addlinespace[5pt]
        \textbf{Random} &
            & \bfseries 37.48 & 19.19 & 24.41 & 18.92 \\
        & & {\footnotesize(26.15--47.65)} & {\footnotesize(11.10--29.46)}
            & {\footnotesize(16.72--34.55)} & {\footnotesize(12.27--25.18)} \\
        \bottomrule
    \end{tabular}

    \caption{\revision{Annotation category composition for TRACE
    (Reference and Frequency selection) versus Random selection.
    Values are percentages of labeled characters
    (95\% confidence intervals), reported separately by annotator.
    The highest percentage in each row is shown in bold.}}
    \label{tab:gold_span_bootstrap}
\end{table}

%% file: SI/FIG/ablation.tex
\begin{table}[htbp]
    \centering
    \small
    \setlength{\tabcolsep}{4pt}
    \renewcommand{\arraystretch}{1.25}
    \begin{tabular}{llcc}
        \hline
        \textbf{Minimum span (characters)} & \textbf{Method} & \textbf{Precision (95\% CI)} & \textbf{Recall (95\% CI)} \\
        \hline
        $\geq$25  & Reference & 0.9568 (0.9303, 0.9747) & 0.8580 (0.7725, 0.9152) \\
        $\geq$25  & TRACE     & 0.8789 (0.8139, 0.9254) & 0.8763 (0.7967, 0.9299) \\
        $\geq$50* & Reference & 0.9614 (0.9343, 0.9792) & 0.8423 (0.7546, 0.9024) \\
        $\geq$50* & TRACE     & 0.8770 (0.8055, 0.9265) & 0.8643 (0.7816, 0.9207) \\
        $\geq$75  & Reference & 0.9623 (0.9330, 0.9812) & 0.7675 (0.6645, 0.8429) \\
        $\geq$75  & TRACE     & 0.8930 (0.8228, 0.9395) & 0.7946 (0.6927, 0.8663) \\
        \hline
    \end{tabular}
    \par\smallskip
    \begin{minipage}{\linewidth}
        \footnotesize *Default threshold used in all other analyses.
    \end{minipage}
    \caption{\revision{Template-detection precision and recall across minimum span thresholds in the 100-note annotated subset. Reference, Reference Module only; TRACE, Reference and Frequency Modules. Recall was evaluated against the same annotated templated spans ($\geq$50 characters) at every threshold. TRACE precision is conservative because templated text flagged only by the Frequency Module is not attribution-linked and therefore counts as a false positive. Values are estimates (95\% confidence intervals from 10{,}000 note-level bootstrap replicates).}}
    \label{tab:ablation_accuracy}
\end{table}

\begin{table}[htbp]
    \centering
    \small
    \setlength{\tabcolsep}{4pt}
    \renewcommand{\arraystretch}{1.25}
    \begin{tabular}{lcc}
        \hline
        \textbf{Minimum span (characters)} & \textbf{Corpus removal (95\% CI)} & \textbf{Timeline removal (95\% CI)} \\
        \hline
        $\geq$25  & 48.97\% (46.53\%, 51.45\%) & 40.58\% (39.73\%, 41.42\%) \\
        $\geq$50* & 47.25\% (44.77\%, 49.76\%) & 38.85\% (38.01\%, 39.67\%) \\
        $\geq$75  & 45.06\% (42.60\%, 47.57\%) & 36.44\% (35.59\%, 37.27\%) \\
        \hline
    \end{tabular}
    \par\smallskip
    \begin{minipage}{\linewidth}
        \footnotesize *Default threshold used in all other analyses.
    \end{minipage}
    \caption{\revision{Text removal across minimum span thresholds in the 1{,}000-patient task-agnostic sample. Corpus removal is the percentage of all characters removed; timeline removal is the mean per-patient percentage removed. Values are estimates (95\% confidence intervals from 10{,}000 patient-level bootstrap replicates).}}
    \label{tab:ablation_removal}
\end{table}

%% file: SI/FIG/notetype_total.tex
\begin{table}[h]
    \centering
    \small
    \begin{tabular}{lccccccc}
        \hline
        \textbf{type} & \textbf{mean} & \textbf{std} & \textbf{min} & \textbf{25} & \textbf{50} & \textbf{75} & \textbf{max} \\
        \hline
        Progress Notes (n=120467) & 0.51 & 0.3 & 0.0031 & 0.23 & 0.5 & 0.8 & 1.0 \\
        Telephone Encounter (n=34780) & 0.36 & 0.31 & 0.0013 & 0.091 & 0.25 & 0.6 & 1.0 \\
        Patient Instructions (n=15399) & 0.72 & 0.28 & 0.0078 & 0.53 & 0.82 & 0.95 & 1.0 \\
        Consults (n=12202) & 0.29 & 0.24 & 0.0039 & 0.1 & 0.21 & 0.4 & 1.0 \\
        H\&P (n=4633) & 0.31 & 0.24 & 0.0039 & 0.13 & 0.25 & 0.44 & 1.0 \\
        ED Provider Notes (n=4156) & 0.2 & 0.15 & 0.0042 & 0.092 & 0.16 & 0.26 & 0.96 \\
        Discharge Instructions (n=3889) & 0.6 & 0.33 & 0.0027 & 0.31 & 0.63 & 0.96 & 1.0 \\
        Procedures (n=3849) & 0.34 & 0.26 & 0.012 & 0.13 & 0.28 & 0.49 & 1.0 \\
        Anesthesia Preprocedure Evaluation (n=2635) & 0.36 & 0.26 & 0.0091 & 0.16 & 0.27 & 0.5 & 1.0 \\
        Discharge Summary (n=2514) & 0.34 & 0.19 & 0.0034 & 0.21 & 0.33 & 0.45 & 1.0 \\
        Anesthesia Postprocedure Evaluation (n=2299) & 0.47 & 0.15 & 0.018 & 0.42 & 0.48 & 0.54 & 1.0 \\
        Assessment \& Plan Note (n=1793) & 0.69 & 0.3 & 0.0087 & 0.44 & 0.77 & 1.0 & 1.0 \\
        Nursing Note (n=1452) & 0.43 & 0.33 & 0.0085 & 0.17 & 0.3 & 0.71 & 1.0 \\
        ED Notes (n=1235) & 0.82 & 0.24 & 0.019 & 0.67 & 0.96 & 1.0 & 1.0 \\
        Anesthesia Procedure Notes (n=1096) & 0.39 & 0.19 & 0.016 & 0.25 & 0.4 & 0.52 & 1.0 \\
        Care Plan (n=1069) & 0.15 & 0.17 & 0.0068 & 0.057 & 0.09 & 0.14 & 1.0 \\
        Unmapped External Note (n=540) & 0.24 & 0.18 & 0.013 & 0.097 & 0.19 & 0.31 & 0.91 \\
        Interval H\&P Note (n=485) & 0.52 & 0.14 & 0.037 & 0.45 & 0.53 & 0.58 & 1.0 \\
        H\&P (View-Only) (n=411) & 0.72 & 0.3 & 0.045 & 0.44 & 0.83 & 1.0 & 1.0 \\
        Subjective \& Objective (n=252) & 0.49 & 0.29 & 0.02 & 0.22 & 0.47 & 0.77 & 1.0 \\
        Addendum Note (n=148) & 1.0 & 0.0 & 1.0 & 1.0 & 1.0 & 1.0 & 1.0 \\
        Group Note (n=109) & 0.31 & 0.25 & 0.028 & 0.15 & 0.2 & 0.39 & 1.0 \\
        Hospice (n=107) & 0.097 & 0.086 & 0.011 & 0.039 & 0.055 & 0.14 & 0.45 \\
        Case Communication (n=86) & 0.2 & 0.22 & 0.019 & 0.039 & 0.11 & 0.32 & 0.89 \\
        Committee Review (n=83) & 0.17 & 0.18 & 0.021 & 0.048 & 0.12 & 0.19 & 1.0 \\
        Lactation Note (n=73) & 0.19 & 0.16 & 0.021 & 0.056 & 0.14 & 0.25 & 0.66 \\
        L\&D Delivery Note (n=29) & 0.41 & 0.33 & 0.023 & 0.19 & 0.29 & 0.54 & 1.0 \\
        Discharge Instr - Other Orders (n=19) & 0.33 & 0.26 & 0.097 & 0.14 & 0.23 & 0.4 & 0.96 \\
        Discharge Instr - Other Info (n=14) & 0.74 & 0.25 & 0.28 & 0.6 & 0.76 & 0.96 & 1.0 \\
        Initial Assessments (n=12) & 0.079 & 0.1 & 0.014 & 0.022 & 0.03 & 0.073 & 0.36 \\
        \hline
    \end{tabular}
    \caption{Annotation by TRACE in 1000 patients for 30 most frequent note types}
    \label{tab:notetype_total}
\end{table}

%% file: SI/FIG/notetype_template.tex
\begin{table}[h]
    \centering
    \small
    \begin{tabular}{lccccccc}
        \hline
        \textbf{type} & \textbf{mean} & \textbf{std} & \textbf{min} & \textbf{25} & \textbf{50} & \textbf{75} & \textbf{max} \\
        \hline
        Progress Notes (n=120467) & 0.16 & 0.22 & 0.0 & 0.023 & 0.075 & 0.2 & 1.0 \\
        Telephone Encounter (n=34780) & 0.27 & 0.29 & 0.0 & 0.061 & 0.13 & 0.4 & 1.0 \\
        Patient Instructions (n=15399) & 0.62 & 0.33 & 0.0 & 0.35 & 0.72 & 0.91 & 1.0 \\
        Consults (n=12202) & 0.14 & 0.15 & 0.0 & 0.045 & 0.094 & 0.18 & 1.0 \\
        H\&P (n=4633) & 0.098 & 0.13 & 0.0 & 0.028 & 0.063 & 0.12 & 1.0 \\
        ED Provider Notes (n=4156) & 0.1 & 0.079 & 0.0 & 0.054 & 0.083 & 0.13 & 0.91 \\
        Discharge Instructions (n=3889) & 0.59 & 0.33 & 0.0 & 0.28 & 0.61 & 0.95 & 1.0 \\
        Procedures (n=3849) & 0.27 & 0.23 & 0.0 & 0.079 & 0.22 & 0.39 & 1.0 \\
        Anesthesia Preprocedure Evaluation (n=2635) & 0.11 & 0.084 & 0.0 & 0.047 & 0.092 & 0.15 & 0.55 \\
        Discharge Summary (n=2514) & 0.12 & 0.099 & 0.0 & 0.061 & 0.11 & 0.17 & 0.93 \\
        Anesthesia Postprocedure Evaluation (n=2299) & 0.46 & 0.14 & 0.0 & 0.41 & 0.47 & 0.53 & 1.0 \\
        Assessment \& Plan Note (n=1793) & 0.092 & 0.24 & 0.0 & 0.0 & 0.0 & 0.0 & 1.0 \\
        Nursing Note (n=1452) & 0.32 & 0.3 & 0.0 & 0.1 & 0.22 & 0.42 & 1.0 \\
        ED Notes (n=1235) & 0.79 & 0.28 & 0.0 & 0.64 & 0.95 & 1.0 & 1.0 \\
        Anesthesia Procedure Notes (n=1096) & 0.37 & 0.2 & 0.0 & 0.22 & 0.38 & 0.51 & 1.0 \\
        Care Plan (n=1069) & 0.082 & 0.082 & 0.0 & 0.046 & 0.078 & 0.1 & 1.0 \\
        Unmapped External Note (n=540) & 0.11 & 0.11 & 0.0 & 0.028 & 0.068 & 0.19 & 0.51 \\
        Interval H\&P Note (n=485) & 0.48 & 0.15 & 0.0042 & 0.44 & 0.52 & 0.56 & 0.98 \\
        H\&P (View-Only) (n=411) & 0.067 & 0.1 & 0.0 & 0.013 & 0.037 & 0.078 & 1.0 \\
        Subjective \& Objective (n=252) & 0.03 & 0.051 & 0.0 & 0.0 & 0.0081 & 0.044 & 0.47 \\
        Addendum Note (n=148) & 0.0 & 0.0 & 0.0 & 0.0 & 0.0 & 0.0 & 0.0 \\
        Group Note (n=109) & 0.19 & 0.16 & 0.0 & 0.12 & 0.17 & 0.21 & 0.97 \\
        Hospice (n=107) & 0.039 & 0.019 & 0.0 & 0.031 & 0.039 & 0.046 & 0.18 \\
        Case Communication (n=86) & 0.013 & 0.036 & 0.0 & 0.0 & 0.0 & 0.0 & 0.19 \\
        Committee Review (n=83) & 0.067 & 0.044 & 0.0 & 0.031 & 0.051 & 0.098 & 0.18 \\
        Lactation Note (n=73) & 0.13 & 0.13 & 0.0 & 0.04 & 0.063 & 0.21 & 0.59 \\
        L\&D Delivery Note (n=29) & 0.16 & 0.16 & 0.0 & 0.023 & 0.17 & 0.22 & 0.6 \\
        Discharge Instr - Other Orders (n=19) & 0.25 & 0.29 & 0.0 & 0.041 & 0.12 & 0.31 & 0.96 \\
        Discharge Instr - Other Info (n=14) & 0.68 & 0.34 & 0.0 & 0.58 & 0.73 & 0.96 & 1.0 \\
        Initial Assessments (n=12) & 0.047 & 0.1 & 0.0 & 0.0 & 0.021 & 0.027 & 0.36 \\
        \hline
    \end{tabular}
    \caption{Template annotation by TRACE in 1000 patients for 30 most frequent note types}
    \label{tab:notetype_template}
\end{table}

%% file: SI/FIG/notetype_copyforward.tex
\begin{table}[h]
    \centering
    \small
    \begin{tabular}{lccccccc}
        \hline
        \textbf{type} & \textbf{mean} & \textbf{std} & \textbf{min} & \textbf{25} & \textbf{50} & \textbf{75} & \textbf{max} \\
        \hline
        Progress Notes (n=120467) & 0.37 & 0.36 & 0.0 & 0.0 & 0.26 & 0.73 & 1.0 \\
        Telephone Encounter (n=34780) & 0.11 & 0.26 & 0.0 & 0.0 & 0.0 & 0.0 & 1.0 \\
        Patient Instructions (n=15399) & 0.13 & 0.3 & 0.0 & 0.0 & 0.0 & 0.0 & 1.0 \\
        Consults (n=12202) & 0.16 & 0.24 & 0.0 & 0.0 & 0.031 & 0.22 & 1.0 \\
        H\&P (n=4633) & 0.22 & 0.25 & 0.0 & 0.021 & 0.14 & 0.35 & 1.0 \\
        ED Provider Notes (n=4156) & 0.095 & 0.14 & 0.0 & 0.0 & 0.031 & 0.13 & 0.96 \\
        Discharge Instructions (n=3889) & 0.027 & 0.13 & 0.0 & 0.0 & 0.0 & 0.0 & 1.0 \\
        Procedures (n=3849) & 0.083 & 0.21 & 0.0 & 0.0 & 0.0 & 0.0 & 1.0 \\
        Anesthesia Preprocedure Evaluation (n=2635) & 0.27 & 0.3 & 0.0 & 0.013 & 0.14 & 0.44 & 1.0 \\
        Discharge Summary (n=2514) & 0.22 & 0.19 & 0.0 & 0.07 & 0.19 & 0.32 & 1.0 \\
        Anesthesia Postprocedure Evaluation (n=2299) & 0.0081 & 0.063 & 0.0 & 0.0 & 0.0 & 0.0 & 1.0 \\
        Assessment \& Plan Note (n=1793) & 0.61 & 0.37 & 0.0 & 0.31 & 0.7 & 1.0 & 1.0 \\
        Nursing Note (n=1452) & 0.12 & 0.29 & 0.0 & 0.0 & 0.0 & 0.0 & 1.0 \\
        ED Notes (n=1235) & 0.033 & 0.17 & 0.0 & 0.0 & 0.0 & 0.0 & 1.0 \\
        Anesthesia Procedure Notes (n=1096) & 0.023 & 0.11 & 0.0 & 0.0 & 0.0 & 0.0 & 1.0 \\
        Care Plan (n=1069) & 0.068 & 0.15 & 0.0 & 0.0 & 0.0 & 0.0 & 1.0 \\
        Unmapped External Note (n=540) & 0.12 & 0.19 & 0.0 & 0.0 & 0.0 & 0.19 & 0.88 \\
        Interval H\&P Note (n=485) & 0.05 & 0.17 & 0.0 & 0.0 & 0.0 & 0.0 & 1.0 \\
        H\&P (View-Only) (n=411) & 0.69 & 0.34 & 0.0 & 0.36 & 0.8 & 1.0 & 1.0 \\
        Subjective \& Objective (n=252) & 0.46 & 0.32 & 0.0 & 0.18 & 0.45 & 0.77 & 1.0 \\
        Addendum Note (n=148) & 1.0 & 0.0 & 1.0 & 1.0 & 1.0 & 1.0 & 1.0 \\
        Group Note (n=109) & 0.14 & 0.26 & 0.0 & 0.0 & 0.0 & 0.1 & 1.0 \\
        Hospice (n=107) & 0.059 & 0.084 & 0.0 & 0.0 & 0.0 & 0.083 & 0.4 \\
        Case Communication (n=86) & 0.19 & 0.23 & 0.0 & 0.028 & 0.1 & 0.32 & 0.89 \\
        Committee Review (n=83) & 0.1 & 0.19 & 0.0 & 0.0 & 0.0 & 0.13 & 0.94 \\
        Lactation Note (n=73) & 0.07 & 0.15 & 0.0 & 0.0 & 0.0 & 0.064 & 0.65 \\
        L\&D Delivery Note (n=29) & 0.26 & 0.4 & 0.0 & 0.0 & 0.0 & 0.33 & 1.0 \\
        Discharge Instr - Other Orders (n=19) & 0.2 & 0.22 & 0.0 & 0.078 & 0.14 & 0.2 & 0.73 \\
        Discharge Instr - Other Info (n=14) & 0.077 & 0.12 & 0.0 & 0.0 & 0.033 & 0.084 & 0.37 \\
        Initial Assessments (n=12) & 0.032 & 0.056 & 0.0 & 0.0 & 0.0 & 0.041 & 0.19 \\
        \hline
    \end{tabular}
    \caption{Copyforward annotation by TRACE in 1000 patients for 30 most frequent note types}
    \label{tab:notetype_copyforward}
\end{table}

%% file: SI/FIG/performance.tex
\begin{table}[h]
    \centering
    \small
    \setlength{\tabcolsep}{3pt}
    \begin{tabular}{llrl}
        \hline
        \textbf{Cohort} & \textbf{Metric} & $\Delta$ & \textbf{95\% CI for $\Delta$} \\
        \hline
        OB/PPH & F1 & $0.0180$ & ($0.0023$, $0.0341$) \\
        OB/PPH & AUROC & $0.0047$ & ($-0.0034$, $0.0127$) \\
        SHC-Inpatient & F1 & $-0.0020$ & ($-0.0141$, $0.0103$) \\
        SHC-Inpatient & AUROC & $0.0003$ & ($-0.0052$, $0.0059$) \\
        MIMIC & F1 & $0.0070$ & ($-0.0106$, $0.0244$) \\
        MIMIC & AUROC & $-0.0033$ & ($-0.0147$, $0.0080$) \\
         Transplant & F1 & $-0.0107$ & ($-0.0325$, $0.0067$) \\

        \hline
    \end{tabular}
    \caption{\revision{Outcome-prediction performance differences with 95\% confidence intervals.}}
    \label{tab:trace-outcome-deltas}
    \par\smallskip

\end{table}

\begin{table}[h]
    \centering
    \small
    \setlength{\tabcolsep}{3pt}
    \begin{tabular}{lp{0.38\linewidth}rl}
        \hline
        \textbf{Cohort} & \textbf{Task} & $\Delta$ & \textbf{95\% CI for $\Delta$} \\
        \hline
        Transplant & Average IE & $0.0040$ & ($-0.0047$, $0.0126$) \\
        Transplant & alcoholic fatty liver disease & $0.0210$ & ($-0.0341$, $0.0784$) \\
        Transplant & ascites & $0.0094$ & ($-0.0100$, $0.0308$) \\
        Transplant & cirrhosis & $-0.0037$ & ($-0.0093$, $0$) \\
        Transplant & cryptogenic cirrhosis & $0$ & ($0$, $0$) \\
        Transplant & hepatic encephalopathy & $0.0091$ & ($0$, $0.0241$) \\
        Transplant & hepatitis b & $0.0158$ & ($0$, $0.0423$) \\
        Transplant & hepatitis c & $-0.0034$ & ($-0.0110$, $0$) \\
        Transplant & hyperlipidemia & $0.0062$ & ($-0.0133$, $0.0282$) \\
        Transplant & hypertension & $0.0083$ & ($-0.0036$, $0.0233$) \\
        Transplant & nonalcoholic fatty liver disease & $-0.0180$ & ($-0.0619$, $0$) \\
        Transplant & obesity & $-0.0101$ & ($-0.0430$, $0.0224$) \\
        Transplant & smoking & $0.0132$ & ($-0.0491$, $0.0762$) \\
        Transplant & variceal bleed & $0.0035$ & ($-0.0356$, $0.0460$) \\
        OB/PPH & Average IE & $-0.0086$ & ($-0.0229$, $0.0043$) \\
        OB/PPH & cerv balloon & $0$ & ($0$, $0$) \\
        OB/PPH & csec urgency & $-0.0300$ & ($-0.0700$, $0$) \\
        OB/PPH & delivery mode & $-0.0300$ & ($-0.0700$, $0$) \\
        OB/PPH & intrapartum & $-0.0300$ & ($-0.0700$, $0$) \\
        OB/PPH & ivf & $0$ & ($0$, $0$) \\
        OB/PPH & labor type & $0.0100$ & ($-0.0300$, $0.0600$) \\
        OB/PPH & num prior csec & $0.0200$ & ($-0.0200$, $0.0600$) \\
        \hline
    \end{tabular}
    \caption{\revision{Aggregate and individual information-extraction F1 differences with 95\% confidence intervals.}}
    \label{tab:trace-ie-f1-deltas}
    \par\smallskip
\end{table}

%% file: SI/FIG/logistic_arm_coefficients.tex
\begin{table}[htbp]
    \centering
    \small
    \setlength{\tabcolsep}{4pt}
    \renewcommand{\arraystretch}{1.2}
    \begin{tabular}{lcccc}
        \toprule
        \textbf{Term} & $\boldsymbol{\beta}$ (SE) & \textbf{OR (95\% CI)} & $\boldsymbol{p}$ & $\boldsymbol{p_{\mathrm{Holm}}}$ \\
        \midrule
        \multicolumn{5}{l}{\textbf{Family 1: Information extraction} \quad (2 hypotheses)} \\
        \addlinespace[3pt]
        \multicolumn{5}{l}{\emph{Transplant: average IE}} \\
        Intercept & $0.863$ ($0.1565$) & $2.37$ ($1.74$, $3.22$) & $3.529\times 10^{-8}$ & --- \\
        TRACE & $0.0496$ ($0.0394$) & $1.05$ ($0.973$, $1.14$) & $0.209$ & $0.4179$ \\
        Prevalence & $2.1238$ ($0.2854$) & $8.36$ ($4.78$, $14.6$) & $9.869\times 10^{-14}$ & --- \\
        \addlinespace[2pt]
        \multicolumn{5}{l}{\emph{Obstetrics: average IE}} \\
        Intercept & $0.4911$ ($0.2606$) & $1.63$ ($0.981$, $2.72$) & $0.05947$ & --- \\
        TRACE & $-0.0731$ ($0.0587$) & $0.929$ ($0.829$, $1.04$) & $0.2124$ & $0.4179$ \\
        Prevalence & $1.9887$ ($0.4391$) & $7.31$ ($3.09$, $17.3$) & $5.913\times 10^{-6}$ & --- \\
        \midrule
        \multicolumn{5}{l}{\textbf{Family 2: Zero-shot outcome prediction} \quad (1 hypothesis)} \\
        \addlinespace[3pt]
        \multicolumn{5}{l}{\emph{Transplant: survival}} \\
        Intercept & $2.8134$ ($0.603$) & $16.7$ ($5.11$, $54.3$) & $3.074\times 10^{-6}$ & --- \\
        TRACE & $-0.5516$ ($0.3931$) & $0.576$ ($0.267$, $1.24$) & $0.1605$ & $0.1605$ \\
        \midrule
        \multicolumn{5}{l}{\textbf{Family 3: Embedding-based outcome prediction} \quad (3 hypotheses)} \\
        \addlinespace[3pt]
        \multicolumn{5}{l}{\emph{Obstetrics: PPH}} \\
        Intercept & $0.3753$ ($0.0673$) & $1.46$ ($1.28$, $1.66$) & $2.493\times 10^{-8}$ & --- \\
        TRACE & $-0.2515$ ($0.0496$) & $0.778$ ($0.706$, $0.857$) & $3.953\times 10^{-7}$ & $\mathbf{7.906\times 10^{-7}}$ \\
        \addlinespace[2pt]
        \multicolumn{5}{l}{\emph{Inpatient-SHC: readmission}} \\
        Intercept & $0.5483$ ($0.0522$) & $1.73$ ($1.56$, $1.92$) & $8.168\times 10^{-26}$ & --- \\
        TRACE & $0.2199$ ($0.0368$) & $1.25$ ($1.16$, $1.34$) & $2.347\times 10^{-9}$ & $\mathbf{7.041\times 10^{-9}}$ \\
        \addlinespace[2pt]
        \multicolumn{5}{l}{\emph{MIMIC-III: readmission}} \\
        Intercept & $-0.4488$ ($0.0975$) & $0.638$ ($0.527$, $0.773$) & $4.139\times 10^{-6}$ & --- \\
        TRACE & $0.2135$ ($0.0732$) & $1.24$ ($1.07$, $1.43$) & $0.003556$ & $\mathbf{0.003556}$ \\
        \bottomrule
    \end{tabular}
    \caption{\revision{Logistic regression coefficients for TRACE versus Original. Endpoint: sensitivity. IE denotes information extraction; prevalence is included only in IE models.}}
    \label{tab:trace-logistic-arm-sensitivity}
    \par\smallskip

\end{table}

\begin{table}[htbp]
    \centering
    \small
    \setlength{\tabcolsep}{4pt}
    \renewcommand{\arraystretch}{1.2}
    \begin{tabular}{lcccc}
        \toprule
        \textbf{Term} & $\boldsymbol{\beta}$ (SE) & \textbf{OR (95\% CI)} & $\boldsymbol{p}$ & $\boldsymbol{p_{\mathrm{Holm}}}$ \\
        \midrule
        \multicolumn{5}{l}{\textbf{Family 1: Information extraction} \quad (2 hypotheses)} \\
        \addlinespace[3pt]
        \multicolumn{5}{l}{\emph{Transplant: average IE}} \\
        Intercept & $-1.0865$ ($0.2572$) & $0.337$ ($0.204$, $0.558$) & $2.392\times 10^{-5}$ & --- \\
        TRACE & $0.0048$ ($0.0291$) & $1$ ($0.949$, $1.06$) & $0.8696$ & $0.8696$ \\
        Prevalence & $4.6439$ ($0.3721$) & $104$ ($50.1$, $216$) & $9.442\times 10^{-36}$ & --- \\
        \addlinespace[2pt]
        \multicolumn{5}{l}{\emph{Obstetrics: average IE}} \\
        Intercept & $0.6199$ ($0.4125$) & $1.86$ ($0.828$, $4.17$) & $0.1329$ & --- \\
        TRACE & $-0.0923$ ($0.0862$) & $0.912$ ($0.77$, $1.08$) & $0.284$ & $0.5681$ \\
        Prevalence & $3.1199$ ($0.5476$) & $22.6$ ($7.74$, $66.2$) & $1.215\times 10^{-8}$ & --- \\
        \midrule
        \multicolumn{5}{l}{\textbf{Family 2: Zero-shot outcome prediction} \quad (1 hypothesis)} \\
        \addlinespace[3pt]
        \multicolumn{5}{l}{\emph{Transplant: survival}} \\
        Intercept & $-1.183$ ($0.1575$) & $0.306$ ($0.225$, $0.417$) & $5.908\times 10^{-14}$ & --- \\
        TRACE & $0.0245$ ($0.1013$) & $1.02$ ($0.84$, $1.25$) & $0.809$ & $0.809$ \\
        \midrule
        \multicolumn{5}{l}{\textbf{Family 3: Embedding-based outcome prediction} \quad (3 hypotheses)} \\
        \addlinespace[3pt]
        \multicolumn{5}{l}{\emph{Obstetrics: PPH}} \\
        Intercept & $1.1237$ ($0.0338$) & $3.08$ ($2.88$, $3.29$) & $2.066\times 10^{-242}$ & --- \\
        TRACE & $0.387$ ($0.0269$) & $1.47$ ($1.4$, $1.55$) & $7.273\times 10^{-47}$ & $\mathbf{2.182\times 10^{-46}}$ \\
        \addlinespace[2pt]
        \multicolumn{5}{l}{\emph{Inpatient-SHC: readmission}} \\
        Intercept & $1.4157$ ($0.0347$) & $4.12$ ($3.85$, $4.41$) & $<1\times 10^{-300}$ & --- \\
        TRACE & $-0.2569$ ($0.0262$) & $0.773$ ($0.735$, $0.814$) & $1.182\times 10^{-22}$ & $\mathbf{2.364\times 10^{-22}}$ \\
        \addlinespace[2pt]
        \multicolumn{5}{l}{\emph{MIMIC-III: readmission}} \\
        Intercept & $1.7162$ ($0.035$) & $5.56$ ($5.19$, $5.96$) & $<1\times 10^{-300}$ & --- \\
        TRACE & $-0.1533$ ($0.0274$) & $0.858$ ($0.813$, $0.905$) & $2.197\times 10^{-8}$ & $\mathbf{2.197\times 10^{-8}}$ \\
        \bottomrule
    \end{tabular}
    \caption{\revision{Logistic regression coefficients for TRACE versus Original. Endpoint: specificity. IE denotes information extraction; prevalence is included only in IE models.}}
    \label{tab:trace-logistic-arm-specificity}
    \par\smallskip
\end{table}

%% file: SI/FIG/logistic_spline_tests.tex
\begin{table}[htbp]
    \centering
    \small
    \setlength{\tabcolsep}{4pt}
    \renewcommand{\arraystretch}{1.2}
    \begin{tabular}{>{\centering\arraybackslash}p{0.32\linewidth} c c c c}
        \toprule
        \textbf{Cohort / task} & \textbf{df} & \shortstack{\textbf{Wald}\\$\boldsymbol{\chi^2}$} & $\boldsymbol{p}$ & $\boldsymbol{p_{\mathrm{Holm}}}$ \\
        \midrule
        \multicolumn{5}{l}{\textbf{Family 1: Information extraction} \quad (2 hypotheses)} \\
        \addlinespace[3pt]
        Transplant: average IE & 4 & $4.1295$ & $0.3888$ & $0.3888$ \\
        Obstetrics: average IE & 4 & $9.5479$ & $0.04877$ & $0.09755$ \\
        \midrule
        \multicolumn{5}{l}{\textbf{Family 2: Zero-shot outcome prediction} \quad (1 hypothesis)} \\
        \addlinespace[3pt]
        Transplant: survival* & 1 & $3.2908$ & $0.06967$ & $0.06967$ \\
        \midrule
        \multicolumn{5}{l}{\textbf{Family 3: Embedding-based outcome prediction} \quad (3 hypotheses)} \\
        \addlinespace[3pt]
        Obstetrics: PPH & 4 & $4.6289$ & $0.3275$ & $0.3275$ \\
        Inpatient-SHC: readmission & 4 & $25.5645$ & $3.873\times 10^{-5}$ & $\mathbf{0.0001162}$ \\
        MIMIC-III: readmission & 4 & $14.2583$ & $0.006515$ & $\mathbf{0.01303}$ \\
        \bottomrule
    \end{tabular}
    \caption{\revision{Wald tests of text-removal associations among TRACE predictions. Endpoint: sensitivity. *Linear term for text removal (1 df) because matrix was singular.}}
    \label{tab:trace-spline-tests-sensitivity}
    \par\smallskip
   
\end{table}

\begin{table}[htbp]
    \centering
    \small
    \setlength{\tabcolsep}{4pt}
    \renewcommand{\arraystretch}{1.2}
    \begin{tabular}{>{\centering\arraybackslash}p{0.32\linewidth} c c c c}
        \toprule
        \textbf{Cohort / task} & \textbf{df} & \shortstack{\textbf{Wald}\\$\boldsymbol{\chi^2}$} & $\boldsymbol{p}$ & $\boldsymbol{p_{\mathrm{Holm}}}$ \\
        \midrule
        \multicolumn{5}{l}{\textbf{Family 1: Information extraction} \quad (2 hypotheses)} \\
        \addlinespace[3pt]
        Transplant: average IE & 4 & $3.1419$ & $0.5344$ & $1$ \\
        Obstetrics: average IE & 4 & $0.7973$ & $0.9388$ & $1$ \\
        \midrule
        \multicolumn{5}{l}{\textbf{Family 2: Zero-shot outcome prediction} \quad (1 hypothesis)} \\
        \addlinespace[3pt]
        Transplant: survival & 4 & $2.506$ & $0.6436$ & $0.6436$ \\
        \midrule
        \multicolumn{5}{l}{\textbf{Family 3: Embedding-based outcome prediction} \quad (3 hypotheses)} \\
        \addlinespace[3pt]
        Obstetrics: PPH & 4 & $8.2893$ & $0.08154$ & $0.08154$ \\
        Inpatient-SHC: readmission & 4 & $123.1925$ & $1.111\times 10^{-25}$ & $\mathbf{2.222\times 10^{-25}}$ \\
        MIMIC-III: readmission & 4 & $129.5579$ & $4.841\times 10^{-27}$ & $\mathbf{1.452\times 10^{-26}}$ \\
        \bottomrule
    \end{tabular}
    \caption{\revision{Wald tests of text-removal associations among TRACE predictions. Endpoint: specificity.}}
    \label{tab:trace-spline-tests-specificity}
    \par\smallskip
  
\end{table}

%% file: SI/FIG/noninferiority_mht.tex
\begin{table}[htbp]
    \centering
    \small
    \setlength{\tabcolsep}{4pt}
    \renewcommand{\arraystretch}{1.2}

    \begin{tabular}{
        >{\centering\arraybackslash}p{0.25\linewidth}
        c c c c c c
    }
        \toprule
        \textbf{Cohort / task}
        & \textbf{Metric}
        & $\boldsymbol{\Delta}$
        & \shortstack{\textbf{Lower}\\\textbf{97.5\% bound}}
        & $\boldsymbol{p}$
        & $\boldsymbol{p_{\mathrm{Holm}}}$
        & \shortstack{\textbf{Non-}\\\textbf{inferior}} \\
        \midrule

        \multicolumn{7}{l}{
            \textbf{Family 1: Information extraction F1}
            \quad (2 hypotheses)
        } \\
        \addlinespace[3pt]

        Transplant: average IE
        & F1 & $0.0040$ & $-0.0046$
        & $4.882\times10^{-35}$
        & $9.764\times10^{-35}$ & Yes \\

        Obstetrics: average IE
        & F1 & $-0.0086$ & $-0.0222$
        & $1.277\times10^{-9}$
        & $1.277\times10^{-9}$ & Yes \\

        \midrule
        \multicolumn{7}{l}{
            \textbf{Family 2: Outcome-prediction AUROC}
            \quad (3 hypotheses)
        } \\
        \addlinespace[3pt]

        Obstetrics: PPH
        & AUROC & $0.0047$ & $-0.0033$
        & $6.451\times10^{-41}$
        & $1.290\times10^{-40}$ & Yes \\

        Inpatient-SHC: readmission
        & AUROC & $0.0003$ & $-0.0053$
        & $6.640\times10^{-71}$
        & $1.992\times10^{-70}$ & Yes \\

        MIMIC-III: readmission
        & AUROC & $-0.0033$ & $-0.0146$
        & $2.757\times10^{-16}$
        & $2.757\times10^{-16}$ & Yes \\

        \midrule
        \multicolumn{7}{l}{
            \textbf{Family 3: Outcome-prediction F1}
            \quad (4 hypotheses)
        } \\
        \addlinespace[3pt]

        Transplant: survival* 
        & F1 & $-0.0107$ & $-0.0302$ 
        & $4.056\times 10^{-5}$ 
        & $4.056\times 10^{-5}$
        & Yes \\

        Obstetrics: PPH
        & F1 & $0.0180$ & $0.0020$
        & $4.297\times10^{-17}$
        & $1.719\times10^{-16}$ & Yes \\

        Inpatient-SHC: readmission
        & F1 & $-0.0020$ & $-0.0140$
        & $3.017\times10^{-15}$
        & $9.050\times10^{-15}$ & Yes \\

        MIMIC-III: readmission
        & F1 & $0.0070$ & $-0.0108$
        & $1.603\times10^{-10}$
        & $3.207\times10^{-10}$ & Yes \\

        \bottomrule
    \end{tabular}

    \caption{\revision{Exploratory Wald non-inferiority tests of TRACE
    versus Original at an absolute margin of 0.05.
    $\Delta$ denotes TRACE minus Original performance.
    For each endpoint, we tested $H_0:\Delta\leq-0.05$
    against $H_1:\Delta>-0.05$.
    Holm correction was applied separately within each
    family shown; non-inferiority was concluded when the
    adjusted one-sided $p$-value was below 0.025.
    Family-wise error was controlled within each family,
    rather than jointly across all nine hypotheses.
    *Survival prediction used zero-shot inference;
    the other outcome-prediction tasks used
    embedding-based classifiers.}}
    \label{tab:trace-noninferiority}
\end{table}

%% file: SI/FIG/out_bootstrap_p.tex
\begin{table}[h]
    \centering
    \small
    \begin{tabular}{lcccc}
        \hline
        \textbf{feature} & \textbf{arm} & \textbf{precision} & \textbf{precision\_lower} & \textbf{precision\_upper} \\
        \hline
        \rowcolor{revisionbg}
        Overall Survival (Liver Transplant) & Original & 0.22 &0.17 & 0.28 \\
        \rowcolor{revisionbg}
        Overall Survival (Liver Transplant) & TRACE & 0.22 & 0.17 & 0.27 \\
        \rowcolor{revisionbg}
        Postpartum Hemorrhage (Obstetrics) & Original & 0.30 &0.28& 0.32\\
        \rowcolor{revisionbg}
        Postpartum Hemorrhage (Obstetrics) & TRACE & 0.34 &0.32& 0.37 \\
        30-Day Readmission (Inpatient-SHC) & Original & 0.52 & 0.49 & 0.54 \\
        30-Day Readmission (Inpatient-SHC) & TRACE & 0.48 & 0.46 & 0.5 \\
        30-Day Readmission (MIMIC-III) & Original & 0.15 & 0.13 & 0.17 \\
        30-Day Readmission (MIMIC-III) & TRACE & 0.15 & 0.13 & 0.17 \\
        \hline
    \end{tabular}
    \caption{Clinical outcome prediction precision with 95\% confidence intervals. OS prediction was evaluated using zero-shot inference.}
    \label{tab:out_bootstrap_p}
\end{table}

%% file: SI/FIG/out_bootstrap_r.tex
\begin{table}[h]
    \centering
    \small
    \begin{tabular}{lcccc}
        \hline
        \textbf{feature} & \textbf{arm} & \textbf{recall} & \textbf{recall\_lower} & \textbf{recall\_upper} \\
        \hline
        \rowcolor{revisionbg}
        Overall Survival (Liver Transplant) & Original & 0.94 & 0.87 & 1.000 \\
        \rowcolor{revisionbg}
        Overall Survival (Liver Transplant) & TRACE & 0.91 & 0.82 & 0.98 \\
        \rowcolor{revisionbg}
        Postpartum Hemorrhage (Obstetrics) & Original & 0.59 &0.56& 0.62 \\
        \rowcolor{revisionbg}
        Postpartum Hemorrhage (Obstetrics) & TRACE & 0.53 &0.50& 0.56 \\
        30-Day Readmission (Inpatient-SHC) & Original & 0.63 & 0.61 & 0.66 \\
        30-Day Readmission (Inpatient-SHC) & TRACE & 0.68 & 0.66 & 0.7 \\
        30-Day Readmission (MIMIC-III) & Original & 0.39 & 0.34 & 0.44 \\
        30-Day Readmission (MIMIC-III) & TRACE & 0.44 & 0.39 & 0.49 \\
        \hline
    \end{tabular}
    \caption{Clinical outcome prediction recall with 95\% confidence intervals. OS prediction was evaluated using zero-shot inference.}
    \label{tab:out_bootstrap_r}
\end{table}

%% file: SI/FIG/out_bootstrap_f.tex
\begin{table}[h]
    \centering
    \small
    \begin{tabular}{lcccc}
        \hline
        \textbf{feature} & \textbf{arm} & \textbf{f1} & \textbf{f1\_lower} & \textbf{f1\_upper} \\
        \hline
        \rowcolor{revisionbg}
        Overall Survival (Liver Transplant) & Original & 0.36 & 0.29 & 0.43 \\
        \rowcolor{revisionbg}
        Overall Survival (Liver Transplant) & TRACE & 0.35 & 0.28 & 0.42 \\
        \rowcolor{revisionbg}
        Postpartum Hemorrhage (Obstetrics) & Original & 0.40 & 0.38 & 0.42 \\
        \rowcolor{revisionbg}
        Postpartum Hemorrhage (Obstetrics) & TRACE & 0.42 & 0.39 & 0.44 \\
        30-Day Readmission (Inpatient-SHC) & Original & 0.57 & 0.55 & 0.59 \\
        30-Day Readmission (Inpatient-SHC) & TRACE & 0.57 & 0.55 & 0.58 \\
        30-Day Readmission (MIMIC-III) & Original & 0.22 & 0.19 & 0.25 \\
        30-Day Readmission (MIMIC-III) & TRACE & 0.23 & 0.20 & 0.25 \\
        \hline
    \end{tabular}
    \caption{Clinical outcome prediction F1 with 95\% confidence intervals. OS prediction was evaluated using zero-shot inference.}
    \label{tab:out_bootstrap_f}
\end{table}

%% file: SI/FIG/out_bootstrap_auroc.tex
\begin{table}[h]
    \centering
    \small
    \begin{tabular}{lcccc}
        \hline
        \textbf{feature} & \textbf{arm} & \textbf{auroc} & \textbf{auroc\_lower} & \textbf{auroc\_upper} \\
        \hline
        \rowcolor{revisionbg}
        Postpartum Hemorrhage (Obstetrics) & Original & 0.747 & 0.729 & 0.763 \\
        \rowcolor{revisionbg}
        Postpartum Hemorrhage (Obstetrics) & TRACE & 0.751 & 0.734 & 0.768 \\
        30-Day Readmission (Inpatient-SHC) & Original & 0.802 & 0.791 & 0.814 \\
        30-Day Readmission (Inpatient-SHC) & TRACE & 0.803 & 0.791 & 0.814 \\
        30-Day Readmission (MIMIC-III) & Original & 0.712 & 0.687 & 0.735 \\
        30-Day Readmission (MIMIC-III) & TRACE & 0.708 & 0.685 & 0.731 \\
        \hline
    \end{tabular}
    \caption{Clinical outcome prediction AUROC with two-sided 95\% percentile-bootstrap confidence intervals from 10,000 paired resampling replicates.}
    \label{tab:out_bootstrap_auroc}
\end{table}

%% file: SI/FIG/outcome_auprc.tex
\begin{table}[h]
    \centering
    \small
    \begin{tabular}{lccc}
        \hline
        \textbf{Outcome} & \textbf{Prevalence} & \textbf{Original} & \textbf{TRACE} \\
        \hline
        \rowcolor{revisionbg}
        Overall Survival (Liver Transplant) & 0.19 & 0.222 & 0.216 \\
        \rowcolor{revisionbg}
        Postpartum Hemorrhage (Obstetrics) & 0.150 & 0.380 & 0.379 \\
        30-day Readmission (Inpatient-SHC) & 0.247 & 0.598 & 0.599 \\
        30-day Readmission (MIMIC-III) & 0.066 & 0.157 & 0.162 \\
        \hline
    \end{tabular}
    \caption{Clinical Outcome Prediction AUPRC}
    \label{tab:outcome_auprc}
\end{table}

%% file: SI/FIG/ablation_pph.tex
\begin{table}[htbp]
    \centering
    \small
    \setlength{\tabcolsep}{6pt}
    \renewcommand{\arraystretch}{1.2}
\def\est#1#2{\shortstack{#1\\{\footnotesize (#2)}}}
\begin{tabular}{lccccc}
        \toprule
\textbf{Input} & \textbf{\shortstack{Minimum span\\(characters)}} & \textbf{AUROC} & \textbf{Precision} & \textbf{Recall} & \textbf{F1} \\        \midrule
        Original & -- & \est{0.747}{0.729, 0.764} & \est{0.299}{0.278, 0.321} & \est{0.593}{0.561, 0.624} & \est{0.398}{0.374, 0.421} \\
        \addlinespace
        TRACE & 25 & \est{0.748}{0.731, 0.765} & \est{0.303}{0.282, 0.323} & \est{0.619}{0.588, 0.650} & \est{0.407}{0.383, 0.429} \\
        \addlinespace
        TRACE & 50\textsuperscript{*} & \est{0.751}{0.734, 0.768} & \est{0.342}{0.317, 0.367} & \est{0.531}{0.499, 0.563} & \est{0.416}{0.391, 0.441} \\
        \addlinespace
        TRACE & 75 & \est{0.745}{0.728, 0.762} & \est{0.341}{0.315, 0.367} & \est{0.480}{0.448, 0.511} & \est{0.399}{0.373, 0.424} \\
        \bottomrule
    \end{tabular}
    \par\smallskip
    {\footnotesize \textsuperscript{*}Default threshold used in all other analyses.\par}
    \caption{\revision{Embedding-based postpartum hemorrhage (PPH) prediction across TRACE minimum span thresholds. Values are estimates (95\% confidence intervals). Original denotes unprocessed notes.}}
    \label{tab:ablation_pph}
\end{table}